\documentclass[journal]{IEEEtran}
\usepackage{cite}
\usepackage{graphicx}
\usepackage{amsmath,amssymb,amsfonts}
\usepackage{algorithm, algorithmic}
\usepackage{subfigure}
\usepackage{multirow}
\usepackage{array}
\usepackage{color}
\usepackage{verbatim}
\begin{document}

\title{Wireless Image Transmission Using Deep Source Channel Coding With Attention Modules}
\author{Jialong~Xu,~\IEEEmembership{Student Member,~IEEE,}
	 Bo Ai,~\IEEEmembership{Fellow,~IEEE,}
        Wei Chen,~\IEEEmembership{Senior Member,~IEEE,}
        Ang Yang,
        Peng Sun,
        Miguel Rodrigues,~\IEEEmembership{Senior Member,~IEEE}% <-this % stops a space
\thanks{This work is supported by the National Key R\&D Program of China (2018YFE0207600, 2020YFB1807201); the Key-Area Research and Development Program of Guangdong Province (2019B010157002); the Natural Science Foundation of China (61911530216, 61961130391, U1834210); the Beijing Natural Science Foundation (L202019); the State Key Laboratory of Rail Traffic Control and Safety (RCS2021ZZ004, RCS2020ZT010) of Beijing Jiaotong University; NSFC Outstanding Youth Foundation under Grant 61725101; the Royal Society Newton Advanced Fellowship under Grant NA191006; vivo research grant.\textit{(corresponding authors: Bo Ai; Wei Chen)}.}
%\thanks{\textit{(corresponding author: Bo Ai, Wei Chen)}.}
\thanks{Jialong Xu is with the State Key Laboratory of Rail Traffic Control and Safety, Beijing Jiaotong University, Beijing 100044, China (e-mail: jialongxu@bjtu.edu.cn).}
\thanks{Bo Ai is with the State Key Laboratory of Rail Traffic Control and Safety, Beijing Jiaotong University, Beijing 100044, China, also with Frontiers Science Center for Smart High-speed Railway System, Beijing 100044, China, also with Research Center of Networks and Communications, Peng Cheng Laboratory, Shenzhen 518055, China, and also with Henan Joint International Research Laboratory of Intelligent Networking and Data Analysis, Zhengzhou University, Zhengzhou 450001, China (e-mail: boai@bjtu.edu.cn).}
\thanks{Wei Chen is with the State Key Laboratory of Rail Traffic Control and Safety, Beijing Jiaotong University, Beijing 100044, China, and also with Frontiers Science Center for Smart High-speed Railway System, Beijing 100044, China (e-mail: weich@bjtu.edu.cn).}
\thanks{Ang Yang and Peng Sun are with vivo Communication Research Institute, Beijing 100015, China (e-mail: ang.yang@vivo.com; sunpeng@vivo.com).}
\thanks{Miguel Rodrigues is with the Department of Electronic and Electrical Engineering, University College London, London, WC1E 7JE,
U.K. (e-mail: m.rodrigues@ucl.ac.uk).}
\thanks{Copyright \copyright 2021 IEEE. Personal use of this material is permitted. However, permission to use this material for any other purposes must be obtained from the IEEE by sending an email to pubs-permissions@ieee.org.}% <-this % stops a space
}


\markboth{IEEE TRANSACTIONS ON CIRCUITS AND SYSTEMS FOR VIDEO TECHNOLOGY}%
{Xu \MakeLowercase{\textit{et al.}}: Attention of SNR in Deep Source Channel Coding}



\maketitle

\begin{abstract}
Recent research on joint source-channel coding (JSCC) for wireless communications has achieved great success owing to the employment of deep learning (DL). However, the existing work on DL based JSCC usually trains the designed network to operate under a specific signal-to-noise ratio (SNR) regime, without taking into account that the SNR level during the deployment stage may differ from that during the training stage. A number of networks are required to cover the scenario with a broad range of SNRs, which is computational inefficiency (in the training stage) and requires large storage. To overcome these drawbacks our paper proposes a novel method called Attention DL based JSCC (ADJSCC) that can successfully operate with different SNR levels during transmission. This design is inspired by the resource assignment strategy in traditional JSCC, which dynamically adjusts the compression ratio in source coding and the channel coding rate according to the channel SNR. This is achieved by resorting to attention mechanisms because these are able to allocate computing resources to more critical tasks. Instead of applying the resource allocation strategy in traditional JSCC, the ADJSCC uses the channel-wise soft attention to scaling features according to SNR conditions. We compare the ADJSCC method with the state-of-the-art DL based JSCC method through extensive experiments to demonstrate its adaptability, robustness and versatility. Compared with the existing methods, the proposed method takes less storage and is more robust in the presence of channel mismatch.

\end{abstract}


% \fi
\IEEEpeerreviewmaketitle

\section{Introduction}
\IEEEPARstart{F}{rom} the first generation to the fifth generation of mobile communication systems,  one traditionally adopts a modular approach to design a communications system (e.g., the transmitter is typically divided into source coder, channel coder and modulation module). Indeed, Shannon's separation theorem \cite{cover1999elements} showcases that a transceiver can be decomposed into a source coding and a channel coding without loss of optimality under certain conditions. However, the theorem assumes that the codeword lengths can be arbitrarily large. This is not realistic in various wireless settings due to latency considerations, such as in emerging wireless enabled applications like autonomous driving, smart manufacturing and telemedicine. 

Shannon's groundbreaking work in 1948\textemdash which asserts that the fundamental problem of communication is that of reproducing at sink either exactly or approximately a message\textemdash does not explicitly dictate the use of separate source-channel coding (SSCC). In fact, Shannon states that if natural redundancy of the source is matched to the statistical characteristics of the channel input, to combat channel noise, there is no need to remove source redundancy \cite{shannon1948}. From then on, joint source-channel coding (JSCC) had become a research hotspot. Gallager gave a mathematical expression of the lower bound of the lossless joint source-channel coding \cite{gallager1968information}.  The expression clearly shows the code length of JSCC is shorter than that of SSCC in the same error performance. Csisz{\'a}r adopted random coding method for the discrete memoryless system containing discrete memoryless source and discrete memoryless channel to give the lower bound and the upper bound expressions of error exponent of JSCC \cite{csiszar1980joint}. Zhong et al. studied error exponents of a point-to-point communication system \cite{zhong2007joint} and a two-user asymmetric communication system \cite{zhong2007error}, and systematically compared the error exponent between SSCC and JSCC. Moreover, the research community has also proposed various concrete JSCC designs. One JSCC strategy leverages SSCC techniques by applying resource assignment \cite{belzer1995joint}, information interaction \cite{heinen2005transactions} and unequal error protection \cite{cai2000robust} to adapt to different signal-to-noise ratios (SNRs). The other JSCC strategy integrates source coding and channel coding as a single process to optimize the communication system \cite{guionnet2004joint}. 

More recently, in view of its impressive performance in domains such as computer vision \cite{szegedy2015going}, speech processing \cite{purwins2019deep} and natural language processing \cite{devlin2018bert}, researchers have also been using deep learning (DL) approaches to support source or channel coding. For example, in the source coding domain, Google research has demonstrated that DL can lead to outstanding image compression results \cite{toderici2015variable,balle2016end,minnen2018joint,minnen2020channel}. Toderici et al. \cite{toderici2015variable} and  Ball{\'e} et al. \cite{balle2016end} demonstrated the DL based autoencoder \cite{goodfellow2016deep} can achieve better compression performance than JPEG and JPEG2000, respectively. Minnen et al. further improved the method of \cite{balle2016end} that completely surpasses the hand-engineered image codec BPG in \cite{minnen2018joint} and \cite{minnen2020channel}. There are also some other works.  To improve the quality of image compression, Jiang et al. \cite{jiang2017end}, Mishra et al. \cite{mishra2020wavelet} and Zhao et al. \cite{zhao2018multiple} proposed to combine existing codecs (e.g., JPEG, JPEG2000 and BPG), wavelet transform and multiple description coding with autoencoders, respectively.

In turn, in the channel coding domain, O’Shea et al. were the first to construct an end-to-end autoencoder with performance close to the Hamming code \cite{o2017introduction}. Considering high-speed mobile channel, Xu et al. introduced measured channel data into the autoencoder, evaluated its performance in real channels and demonstrated the advantage of the improved autoencoder in channel mismatch condition \cite{xu2019performance}. Jiang et al. proposed Low-latency Efficient Adaptive Robust Neural in \cite{jiang2020learn}  and TurboAE code in \cite{jiang2019turbo} to demonstrate the superiority of channel coding based on DL in comparison with conventional channel coding techniques, in the regime of short-medium block-length. Other approaches adopted to support effective wireless communications have resorted to reinforcement learning and generative adversarial networks \cite{aoudia2018end,ye2018channel}.

These successes have in turn also motivate the use of DL based JSCC. For example, \cite{farsad2018deep} has proposed DL based JSCC for the transmission of text over wireless communications channels for the first time. \cite{bourtsoulatze2019deep} designed a joint source-channel encoder and a joint source-channel decoder for image source. This DL based JSCC scheme outperforms SSCC schemes combining JPEG or JPEG2000 with capacity-achieving channel codes. \cite{kurka2019successive} proposed three hierarchical DL based JSCC schemes leveraging the joint source-channel encoder and joint source-channel decoder of \cite{bourtsoulatze2019deep} as basic structures for successive refinement of images. \cite{kurka2020deepjscc} proposed a DL based JSCC scheme that incorporates channel output symbols to the transmission system, which further improves the performance of DL based JSCC. 
Considering a discrete channel, \cite{choi2019neural} proposed a new discrete variational autoencoder model named Neural Error Correcting and Source Trimming based on the maximization of the mutual information between the source and noisy received codeword. Variational inference for Monte Carlo objectives \cite{mnih2016variational} is used to circumvent the non-differential network introduced by the discrete channel. For processed sources, \cite{jankowski2020deep} used the joint source-channel encoder for dimension reduction and the joint source-channel decoder for feature recovery to transmit the person's features extracted by the neural network based on ResNet-50 architecture \cite{he2016deep}. For the extension of  JSCC, \cite{lee2019deep} proposed a joint transmission-recognition scheme that builds the encoder combining the feature extraction with the joint source-channel encoder and the decoder combining recognition with the joint source-channel decoder to transmit the encoded feature wirelessly to the server for recognition tasks. 

\begin{figure}[t]
\centering
\includegraphics[width=1\linewidth]{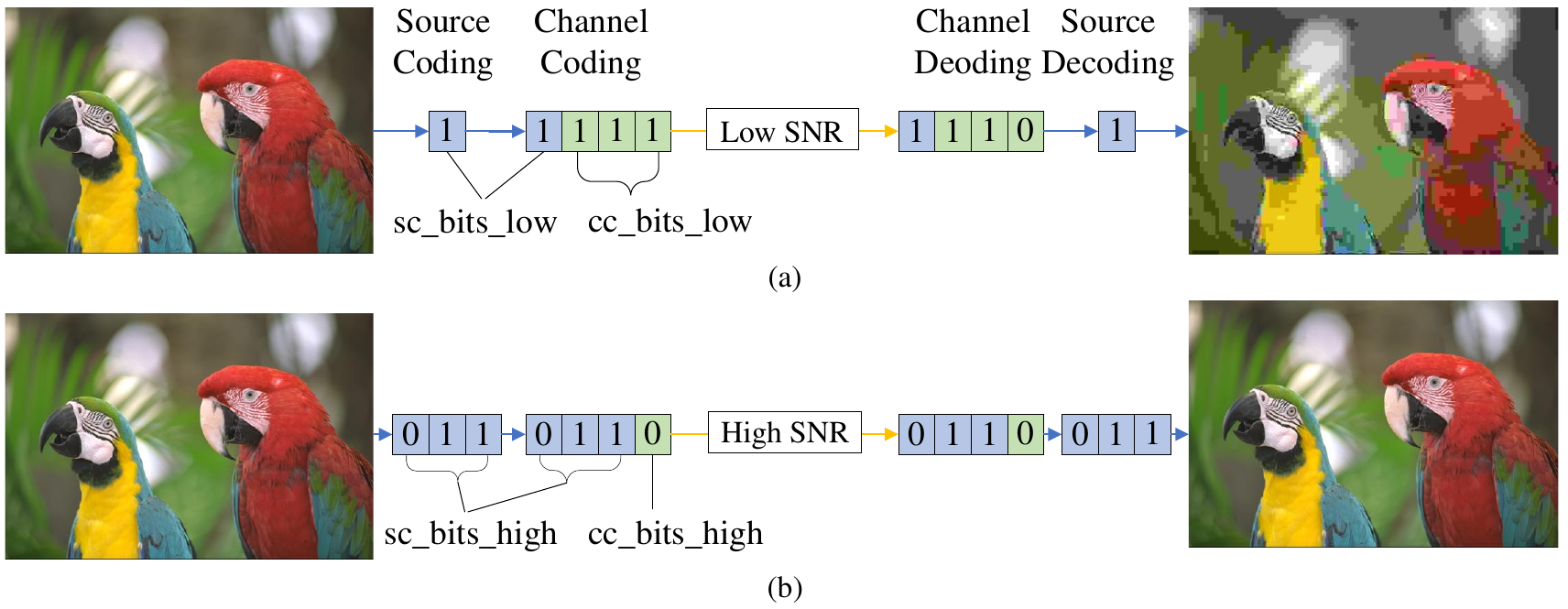}
\caption{Illustration of traditional JSCC. (a) In the low SNR regime, more bits (cc\_bits\_low) are allocated for channel coding and fewer bits (sc\_bits\_low) are allocated for source coding, (b) In the high SNR regime, fewer bits (cc\_bits\_high) are allocated for channel coding and more bits (sc\_bits\_high) are allocated for source coding.}\label{Fig:jscc}
\end{figure}

However, the existing DL based JSCC methods assume that the channel conditions\textemdash notably the SNR\textemdash used to optimize the networks are the same as the channel conditions experienced during network deployment in the actual communications system. However, as shown in \cite{bourtsoulatze2019deep} and \cite{burth2020joint}, a mismatch between the channel conditions used during the network optimization and network deployment stage can lead to serious performance degradation, limiting the advantages of DL based JSCC techniques. In principle, one could train multiple networks suited to a range of SNR that would then be selected during transmission/reception depending on the exact SNR condition, but this would lead to transceivers exhibiting considerable computational/storage requirements limiting its applicability in resource-constrained application such as IoT. In summary, in view of the fact that wireless channels can experience varying SNR levels, one needs to develop new DL based JSCC methods capable of adapting to varying channel conditions. 
In this paper, we design a new DL based JSCC method\textemdash leveraging traditional JSCC design principles\textemdash that can operate successfully over a wide range of SNRs. The inspiration of our method originates from the resource assignment strategy adopted by traditional JSCC illustrated in Fig.~\ref{Fig:jscc}. Specifically, when the channel exhibits bad condition, for the same image, more bits are allocated to the channel encoder and fewer bits are allocated to the source encoder. The increased number of bits allocated to the channel encoder improve the redundancy to combat the intense channel noise. Conversely, when the channel is in good condition, for the same image, fewer bits are given to the channel encoder and more bits are given to the source encoder. The increased source bits are used to improve image quality. Some cross-layer optimization for image/video streaming approaches drawing on traditional concatenated source channel coders have demonstrated to be effective as shown in \cite{loiacono2010cross, zhao2016ssim}. In our proposed method,  channel-wise soft attention network is used to replace the artificially designed resource allocation strategy to dynamically adjust the compression ratio in source coding and the channel coding rate according to the range of SNR. Compared with \cite{bourtsoulatze2019deep} where neither the transmitter nor the receiver has channel SNR knowledge, our proposed method leads to a large performance improvement. Compared with \cite{kurka2020deepjscc} where one has only access to delayed SNR knowledge based on feedback from receiver to transmitter, our proposed method uses attention mechanism as a resource allocation approach to allocate different contributions for intermediate features in the coding process according to channel SNR;  it also feedback channel information more efficiently because it relays channel SNR back to transmitter rather than the channel output.
Another advantage of our proposed method is that our proposed method is more robust than the method proposed in \cite{bourtsoulatze2019deep} in the presence of channel mismatch. The most important contribution of our work is that we have successfully explored a way to design the JSCC base DL method with the help of traditional JSCC design principles in the development of wireless communications. 

The rest of this work is organized as follows. In Section II, we introduce the system model. Then, the proposed method is presented in Section III. In Section IV, we offer a series of simulation results to showcase the performance of our method. Section V is dedicated to the evaluation of the performance and the storage of the proposed method compared with other DL based JSCC schemes. Finally, the paper is concluded in Section VI.

\section{System Model}

\begin{figure}[!htb]
\centering
\includegraphics[width=1\linewidth]{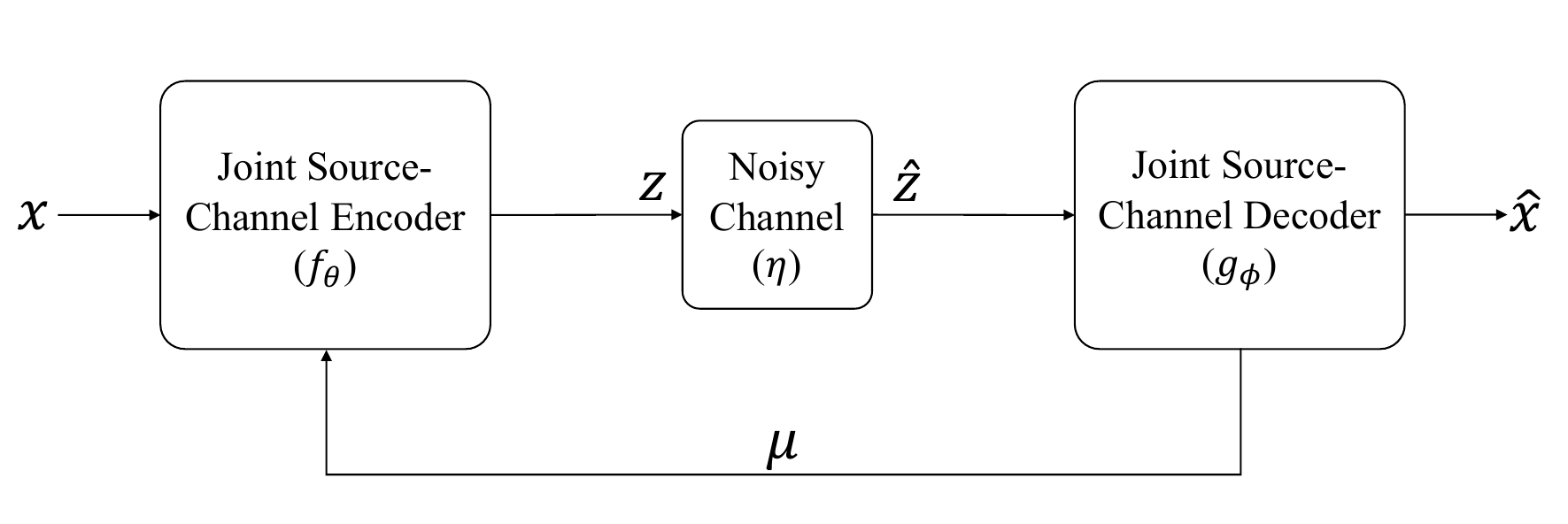}
\caption{The model of the point-to-point image transmission system with SNR feedback.}\label{Fig:snr_feedback}
\end{figure}

Consider a point-to-point image transmission system with SNR feedback as shown in Fig.~\ref{Fig:snr_feedback}. Channel SNR is known both at the joint source-channel encoder and the joint source-channel decoder. An input image of size H(height)$\times$W(weight)$\times$C(channel) is represented by a vector $\boldsymbol{x} \in \mathbb{R}^n$, where $n$ = H$\times$W$\times$C and $\mathbb{R}$ denotes the set of real numbers. The joint source-channel encoder encodes $\boldsymbol{x}$ and the feedback SNR $\mu$, and the encoding function $f_{\theta}: \mathbb{R}^n \times \mathbb{R} \rightarrow \mathbb{C}^k$ leads to a vector of complex-valued channel input symbols $\boldsymbol{z} \in \mathbb{C}^k$. The encoding process can be expressed as:
\begin{equation} 
\boldsymbol{z}=f_\theta(\boldsymbol{x}, \mu) \in \mathbb{C}^k,
\label{encode}
\end{equation}
where $k$ is the size of channel input symbols, $\theta$ is the parameter set of the joint source-channel encoder, $\mu \in \mathbb{R}$ is the channel SNR that can be estimated at the joint source-channel decoder and fed back to the joint source-channel encoder, and $\mathbb{C}$ denotes the set of complex numbers. The encoder maps the n-dimensional vector of real-valued image $\boldsymbol{x}$ to a k-dimensional vector of complex-valued channel input samples $\boldsymbol{z}$. To satisfy the average power constraint at the joint source-channel encoder, $ \frac{1}{k}\mathbb{E}(\boldsymbol{zz^*})\leq 1$ is also imposed, where $\boldsymbol{z^*}$ denotes the conjugate transpose of $\boldsymbol{z}$. 

The encoded symbols $\boldsymbol{z}$ are transmitted over a noisy channel represented by the function $\eta: \mathbb{C}^k \rightarrow \mathbb{C}^k$. AWGN channel is considered in our work.
The channel output symbols  $\boldsymbol{\hat{z}} \in \mathbb{C}^k$ received by the joint source-channel decoder are expressed as: 
\begin{equation} 
\boldsymbol{\hat{z}}=\eta(\boldsymbol{z})=\boldsymbol{z} + \boldsymbol{\omega},  
\label{channel}
\end{equation}
where the vector $\boldsymbol{\omega} \in \mathbb{C}^k$ consists of independent and identically distributed (i.i.d) samples with the distribution $\mathbb{CN}(0, \sigma^2\boldsymbol{I})$. $\sigma^2$ is the noise power and $\mathbb{CN}(\cdot,\cdot)$ denotes a circularly symmetric complex Gaussian distribution. The proposed method can be applied to other differentiable channels. Consider the following fading channel:
\begin{equation} 
\boldsymbol{\hat{z}}=h\boldsymbol{z} + \boldsymbol{\omega},
\label{fading_channel}
\end{equation}
where $h \in \mathbb{C}$ is the channel gain. By applying equalization at the receiver, the above model can be represented as Eq. \eqref{channel}, while the noise has a different distribution. The study on non-differentiable channel is out of the scope of this paper.

The joint source-channel decoder uses a decoding function $g_\phi: \mathbb{C}^k \times \mathbb{R} \rightarrow \mathbb{R}^n$ to map $\boldsymbol{\hat{z}}$ and $\mu$ as follows:

\begin{equation} 
\boldsymbol{\hat{x}}=g_\phi(\boldsymbol{\hat{z}}, \mu)=g_\phi(\eta(f_\theta(\boldsymbol{x}, \mu)),\mu),
\label{decode}
\end{equation}
where $\boldsymbol{\hat{x}} \in \mathbb{R}^n$ is an estimation of the original image $\boldsymbol{x}$, $\phi$ is the parameter set of the joint source-channel decoder. The distortion between the original image $\boldsymbol{x}$ and the reconstructed image $\boldsymbol{\hat{x}}$ is expressed as: 
\begin{equation} 
d(\boldsymbol{x},\boldsymbol{\hat{x}})=\frac{1}{n}\sum_{i=1}^n(x_i-\hat{x}_i)^2,
\label{single_distortion}
\end{equation}
where $x_i$ and $\hat{x}_i$ represents the intensity of the color component of each pixel corresponding to $\boldsymbol{x}$ and $\boldsymbol{\hat{x}}$, respectively.

Akin to \cite{bourtsoulatze2019deep,kurka2019successive,kurka2020deepjscc}, we call the image size $n$, the channel input size $k$ and $R=k/n$ as the source bandwidth, the channel bandwidth and bandwidth ratio, respectively. Under a certain $R$, the goal is to determine the encoder and decoder parameters $\theta^*$ and $\phi^*$ that minimize the expected distortion as follows:
\begin{equation} 
(\theta^*, \phi^*)=\mathop{\arg\min}\limits_{\theta, \phi}\mathbb{E}_{p(\mu)}\mathbb{E}_{p(\boldsymbol{x},\boldsymbol{\hat{x}})}[d(\boldsymbol{x},\boldsymbol{\hat{x}})],
\label{argmin}
\end{equation}

where $\theta^*$ is the optimal encoder parameters, $\phi^*$ is the optimal decoder parameters, $p(\boldsymbol{x},\boldsymbol{\hat{x}})$ represents the joint probability distribution of the original image $\boldsymbol{x}$ and the reconstructed image $\boldsymbol{\hat{x}}$, and $p(\mu)$ represents the probability distribution of the SNR. We will model the encoder and decoders using deep neural networks.

\section{proposed method}
\label{section3}

The majority of existing DL based JSCC approaches are designed to operate under specific SNR \cite{bourtsoulatze2019deep,kurka2019successive,kurka2020deepjscc,jankowski2020deep,lee2019deep}.
There are also some recent JSCC based DL techniques that operate under a range of SNRs but these involve optimizing a series of networks for specific SNRs during the training stage and using a specific network adequate for current SNR conditions during the testing stage. This leads to serious drawbacks, including higher computational requirements during the training stage and higher storage demands during the testing stage.

Our goal is to design a single network for joint source-channel coding that can adapt to a wide range of SNR conditions. The proposed method is motivated by the resource assignment strategy in traditional concatenated source channel coders \cite{sayood2000joint} which concatenate the source encoder and the channel encoder, and adjust the compression ratio and the channel coding rate according to the SNR to achieve the optimal quality of reconstructed images under the limited bandwidth. To transmit an image with some fixed bandwidth resource in the low SNR regime, one compresses the source more aggressively but simultaneously increases the channel coding rate in order to combat channel induced errors. On the other hand, in the high SNR regime, one compresses the source less aggressively but decreases the channel coding rate. This approach allows for nearly optimal transmission under a constant rate. Unfortunately, the existing DL based JSCC methods do not support a flexible network structure that can automatically change and adapt to the channel state. 

We address this challenge by employing attention mechanism, which is a technique of DL widely used in natural language processing \cite{bahdanau2014neural,luong2015effective,vaswani2017attention} and computer vision \cite{mnih2014recurrent,wang2017residual,hu2018squeeze,woo2018cbam}. Such mechanisms use an additional neural network that can rigidly select certain features or assign different weights to different features in the original neural network. Our approach\textemdash so-called Attention DL based JSCC (ADJSCC)\textemdash has been constructed to specifically implement joint source-channel coding.

\begin{figure*}[!htb]
\centering
\includegraphics[width=2\columnwidth]{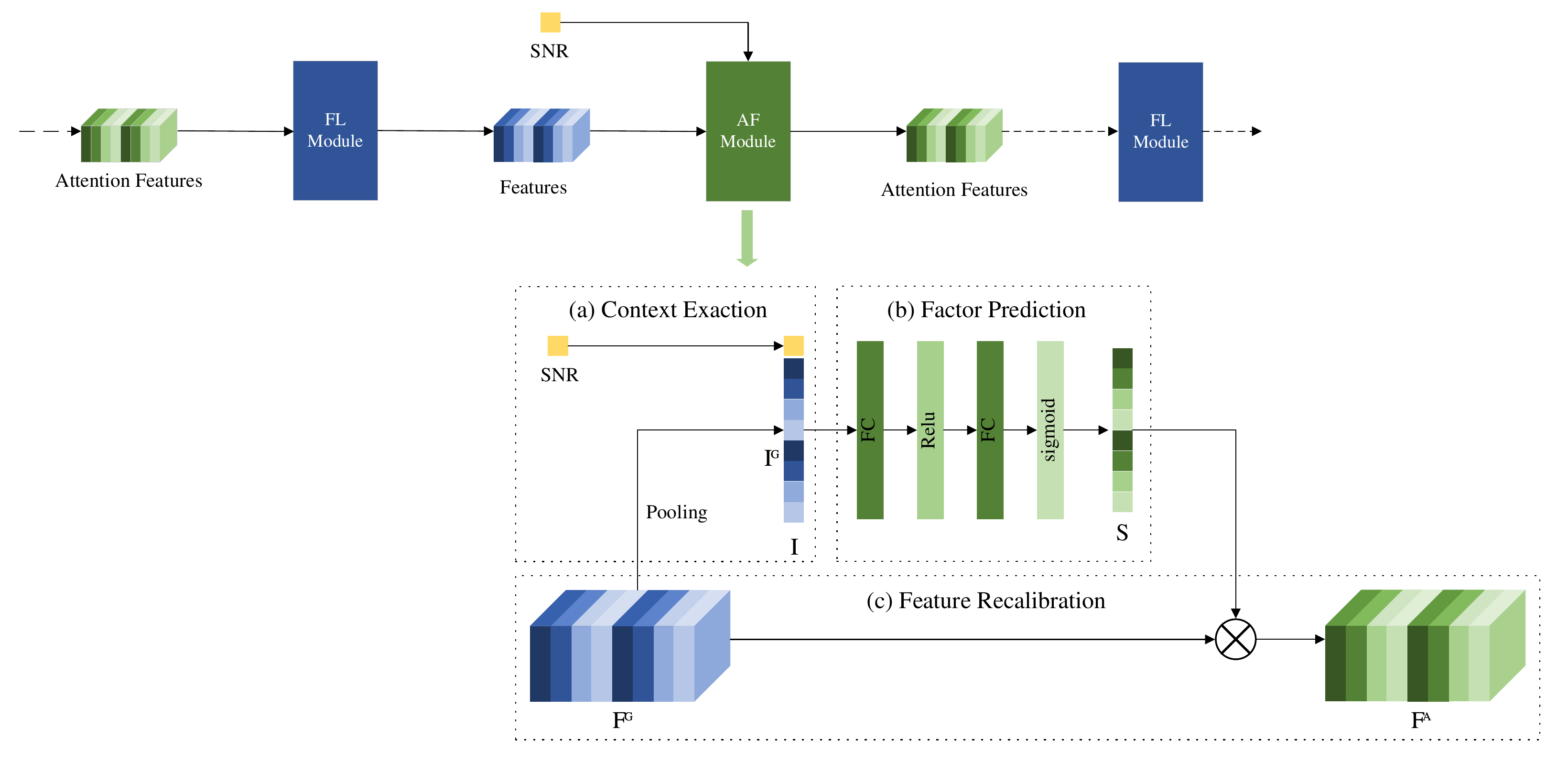}
\caption{The relationship between FL modules and AF modules in ADJSCC. FL Module based on convolution layers has been studied in several works [29]–[32]. AF module contains three parts: (a) Context Extraction, (b) Factor Prediction, (c) Feature Recalibration.}\label{Fig:fl_af_module}
\end{figure*}

The architecture of ADJSCC has two parts, i.e., the neural encoder at the transmitter and the neural decoder at the receiver. The neural encoder usually consists of multiple non-linear layers. The first few layers of the neural encoder can be seen as the source encoder, and the remaining layers of the neural encoder are regarded as the channel encoder. The input and output of the neural encoder in our proposed method are source values and channel symbols, respectively. Thus, it performs the source encoder function and the channel encoder function. The neural decoder in turn perform the reverse operations. 
Our approach exhibits two additional features: (a) first, one can adapt the compression rate as a function of the SNR conditions by allowing the source encoder to output more or less symbols; (b) second, one can also dynamically adjust the size of the sub-networks associated with the source encoder and channel encoder using attention mechanisms, which is akin to the adaptation of the source compression ratio and the channel coding rate according to the SNR.

We now describe in detail the proposed ADJCC design. Both the proposed encoder and decoder are constructed using two types of modules, i.e., a feature learning (FL) module and an attention feature (AF) module. FL modules and AF modules are connected alternately as shown in the upper part of Fig.~\ref{Fig:fl_af_module}. FL module learns features from the input of the FL module, and AF module takes the SNR and the output of the FL module as the input and produces a sequence of scaling parameters. The product of the outputs of the FL module and the AF module can be seen as a filtered version of the FL module output. The design of the FL module based on convolution layers has been studied in several works \cite{bourtsoulatze2019deep,kurka2019successive,kurka2020deepjscc,choi2019neural}. Therefore we focus on the design of AF module.

The ideal hard attention mechanism generates a mask containing elements equal to either 0 or 1 that changes the size of the effective (i.e., nonzero) features. However, the non-differentiable property of the loss in hard attention hinders the execution of the backpropagation algorithm in the training stage. Instead, soft attention is generally adopted instead of hard attention to facilitate backpropagation. The extracted features can be regarded as the signal components on the convolution kernel. The channel relationship are captured and different scaling parameters are generated for different channel features to increase or suppress their connection strength to the next layer. The aforementioned mechanism is channel-wise soft attention\footnote{The notion channel in ``channel-wise soft attention'' refers to the feature channel rather than the communications channel.}.

The architecture of AF module based on channel-wise soft attention is shown in the lower part of Fig.~\ref{Fig:fl_af_module}. Let $F^G=[F^G_1, F^G_2, \cdots, F^G_c] \in \mathbb{R}^{h \times w \times c}$ denote the features extracted by the FL module, where $c$ is the number of the features and $h \times w$ is the size of each feature. Let also $F_{A}=[F^A_1, F^A_2, \cdots F^A_c] \in \mathbb{R}^{h \times w \times c}$ denote the scaled features produced by AF module. In summary, the AF module processes the features $F^G$ using a global average pooling function. The results of the global average pooling is then concatenated with SNR to form the context information. The context information is fed into a full connected neural network that produces a scaling factor. The scaled features $F^{A}$ are obtained by multiplying the features $F^G$ and the scaling factor. This  process results in different scaled features $F^{A}$ depending on the exact SNR conditions.

We next elaborate further about the different components of the AF module: 1) context extraction; 2) factor prediction; and 3) feature recalibration.

\subsubsection{Context Extraction}
Context information includes channel SNR $\mu$ and feature information $I^G$. Image features are usually extracted by convolution kernels limited in a local receptive field. Hence these features cannot usually perceive the information out of this region especially using feature extraction with small kernel size. Global average pooling $G(\cdot)$ however can extract global information by averaging elements $u_{jk}$ of $F^G_i$ as follow:
\begin{equation} 
I^G_i = G(F^G_i) = \frac{1}{h\times w}\sum_{j=1}^h\sum_{k=1}^wu_{jk} \in \mathbb{R}. 
\label{CF}
\end{equation}
These global information features $I^G_i$ are concatenated with the channel information SNR = $\mu$ to produce the context information $I$ as follows: 
\begin{equation} 
I=(\mu, I^G_1, I^G_2, \cdots, I^G_c) \in \mathbb{R}^{c+1}.
\label{i}
\end{equation}
%\textcolor[RGB]{79,148,205}{text}
\subsubsection{Factor Prediction}
We employ a factor prediction neural network $P_\omega(\cdot)$ to predict the scaling factor $S$ based on the context information $I$. In order not to excessively increase the limit complexity, $P_\omega(\cdot)$ is a simple neural network consisting of two fully connected (FC) layers. The first FC layer contains a ReLu and the last FC layer contains a sigmoid in order to limit the output range to the interval (0,1). Therefore,
\begin{equation} 
S = P_\omega(I) = \sigma(W_2\delta(W_1I+b_1)+b_2)\in \mathbb{R}^c, 
\label{FP}
\end{equation}
where $W_1$ and $b_1$ refer to the weights and the biases of the first FC layer, $W_2$ and $b_2$ refer to the weights and the biases of the second FC layer, and $\delta$ and $\sigma$ represent the activation function ReLu and sigmoid, respectively. We let $\omega$ = ($W_1$, $b_1$, $W_2$, $b_2$) denote the parameter set of the  factor prediction neural network.

\subsubsection{Feature Recalibration} Finally, feature recalibration produces a feature map $F^A$ based on the feature map $F^G$ as follows:
\begin{equation} 
F^A_i = R_e(F^G_i, S_i) = S_i \cdot F^G_i, i=1,2,\cdots,c,
\label{FR}
\end{equation}
where $F_i^A$ denotes the i-th element in $F^A$, $F_i^G$ denotes the i-th element in $F^G$, and $S_i$ denotes the i-th element of $S$. The operation of our AF module is described in Algorithm \ref{alg:1}. 
%The work of \cite{hu2018squeeze} is more close to our work. They use image inherent features extracted by the neural network to execute global average-pooling then compute channel-wise attention. 
Different from the attention mechanism used in computer vision, we use SNR as the wireless channel information combined with the image inherent features to compute channel-wise attention. Once again, our motivation derives from the fact that the number of bits that one ought to allocate to source coding and channel coding should depend on the channel SNR. 
%Discussing the effect of the number of the neurons of the first layer in the K-net is out of our scope. Here we follow the \cite{hu2018squeeze} to set this number to the input number of the K-net divided by 16.

\begin{algorithm}
	\renewcommand{\algorithmicrequire}{\textbf{Input:}}
	\renewcommand{\algorithmicensure}{\textbf{Output:}}
	\caption{AF module}
	\label{alg:1}
	\begin{algorithmic}[1]
		\REQUIRE the features $F^G$, the SNR information $\mu$		
		\ENSURE the attention features $F^A$
		\STATE calculate the size of the features: $(height, width, channel) = {\rm size}(F^G)$ 
		\STATE calculate the global information features: $I^G = {\rm GlobalAveragePooling}(F^G)$
		\STATE calculate the context information $I = {\rm concatenate}(\mu, I^G)$
		\STATE calculate the scaling factor $S = P_\omega(I)$
		\STATE convert the scaling factor $S$ to the channel-wise scaling factor $S_i,  i=1, 2, \cdots, c$
		\STATE convert the features $F^G$ to the channel-wise feature $F^G_i,  i=1, 2, \cdots, c$
		\FOR{i = 0 : 1 : c}
		\STATE channel-wise attention feature: $F^A_i = S_i \cdot F^G_i$
		\ENDFOR
		\STATE convert the channel-wise attention feature $F^A_i, i=1, 2, \cdots, c$ to the attention features $F^A$
		\STATE \textbf{return} $F^{A}$
	\end{algorithmic}  
\end{algorithm}

%The other type of module in our ADJSCC is the GFR Module. The GFR Module execute a series of convolution, stride, pooling, activation operations to transform the input image or previous layer feature maps into the output feature maps. Most of the layers in the DJSCC methods \cite{bourtsoulatze2019deep} \cite{kurka2019successive} \cite{kurka2020deepjscc} could be selected as our GFR Module. The proposed AF Module could be applied to these layers by simply inserting the AF Module into the layer which we regard as GFR Module and introducing the channel information into the AF Module.

\section{Simulation Results}
\label{s4}

\begin{figure*}[!htb]
\centering
\includegraphics[width=2\columnwidth]{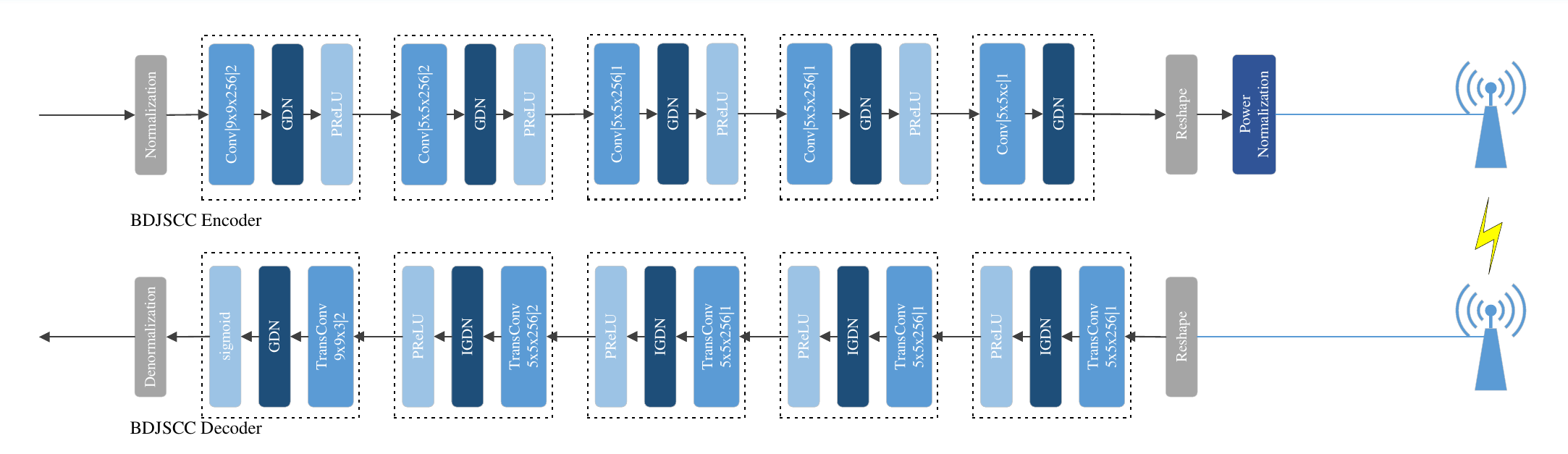}
\caption{The architecture of BDJSCC used in \cite{kurka2020deepjscc}. Convolution and transposed convolution are parameterized by $F \times F \times K | S$, where $F$ and $K$ are the filter size and the number of filters, respectively. In the convolution layer, $S$ represents downsampling strides. In the transposed convolution layer,  $S$ represents upsampling strides.}\label{Fig:bdjscc}
\end{figure*}

\begin{figure*}[!htb]
\centering
\includegraphics[width=2\columnwidth]{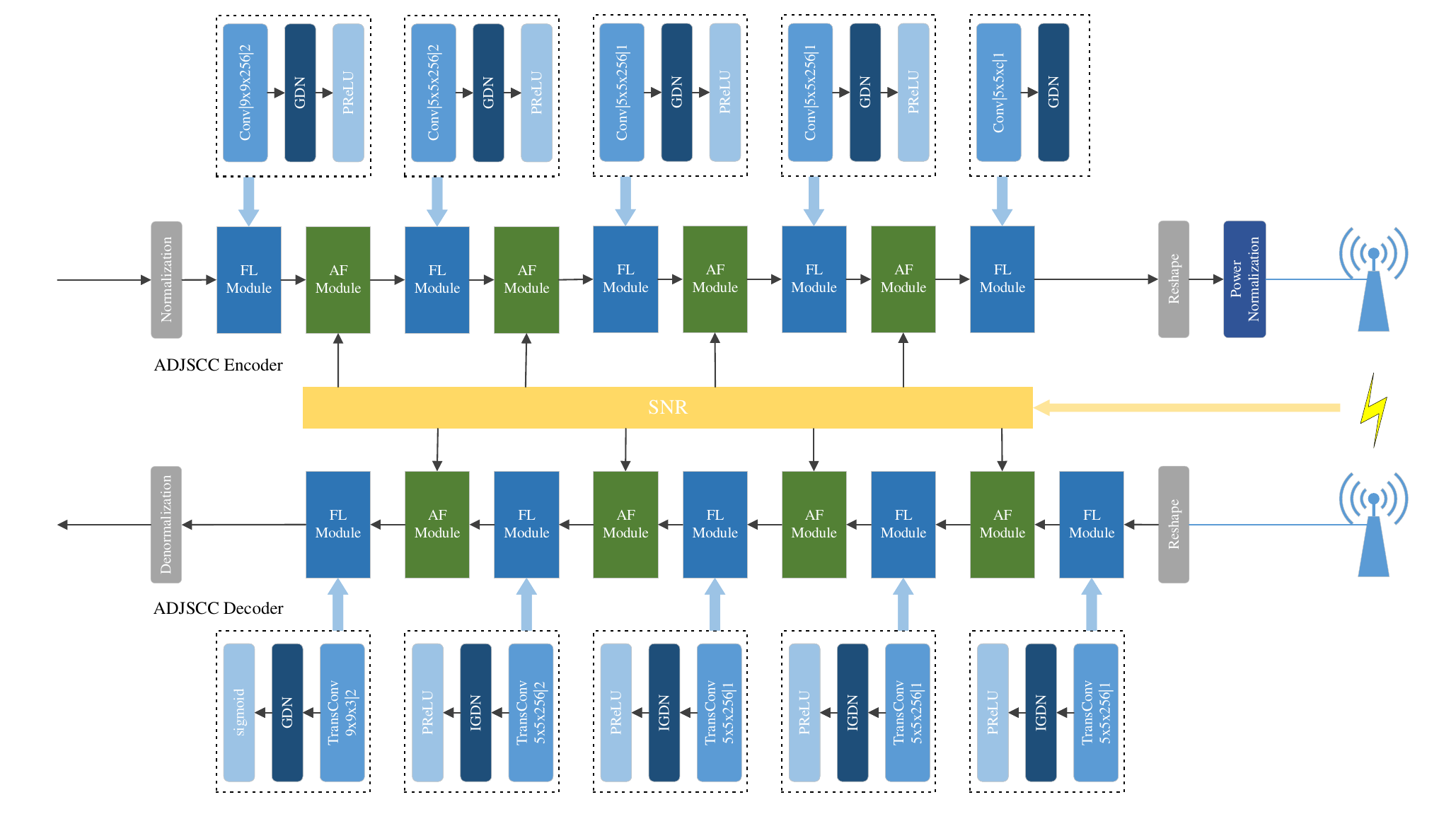}
\caption{The architecture of our proposed ADJSCC. The FL modules of ADJSCC consists of a convolution layer, a GDN layer and a PReLU (or sigmoid) layer. Each FL module is followed by an AF module except the last FL module in the encoder and the decoder. The SNR coming from channel feedback is another input of the AF module.}\label{Fig:adjscc}
\end{figure*}

The first DL based JSCC architecture proposed by \cite{bourtsoulatze2019deep} consists of convolutional modules for image source. Each module consists of a convolution layer followed by a parametric ReLU (PReLU) every layer except the last one or a sigmoid activation function in the last layer. It has led to comparable performance to the standard SSCC scheme (JPEG/JPEG2000 + LDPC). \cite{kurka2020deepjscc} further improved the performance by introducing the generalized divisive normalization (GDN) as a normalization method and widening the channel of the convolution layer for each module. To prove the efficiency of our proposed ADJSCC scheme, we adopt the state-of-the-art DL based JSCC architecture used in \cite{kurka2020deepjscc} as the basic DL based JSCC (BDJSCC) architecture as shown in Fig.~\ref{Fig:bdjscc}. The layers of the BDJSCC Encoder except the normalization layer, the reshape layer and the power normalization layer are divided into five modules. Each of the first four modules consists of a convolution layer, a GDN layer \cite{balle2015density} and a PReLU layer \cite{he2015delving}. The fifth module only consists of a convolution layer and a GDN layer. Similarly, the layers of the BDJSCC Decoder except for the normalization layer and the reshape layer are also divided into five modules. The first four modules  have the same construction consisting of a transposed convolution layer, a GDN layer and a PReLU layer. The only difference between the last module and the first four modules is that the sigmoid layer replaces the PReLU layer. The notation $F \times F \times K | S$ in a convolution/transposed convolution layer denotes that it has $K$ filters with size $F$ and stride down/up $S$. 
%We compare our approach exclusively to BDJSCC that is known to deliver state-of-the-art image transmission performance.

The corresponding ADJSCC architecture\footnote{Source codes for constructing the ADJSCC architecture and the BDJSCC architecture are available at: https://github.com/alexxu1988/ADJSCC.} is shown in Fig.~\ref{Fig:adjscc}. 
%FL Modules of ADJSCC consist of a convolution layer, a GDN layer and a PReLU (or sigmoid) layer and AF modules adopt the method proposed in Fig.~\ref{Fig:adjscc} shows the method that extends a BDJSCC architecture to an ADJSCC architecture. 
Five modules in BDJSCC Encoder are considered as five FL modules, each of which is followed by an AF module except the last FL module. The output of the previous FL module is one input of the present AF module and the output of the present AF module is the input of the next FL module. The other input of the AF module is the SNR level associated with the communications channel. The ADJSCC Decoder is also constructed similarly. By changing the output channel size $c$ in the last convolution layer of the encoder, different bandwidth ratios can be obtained. 

To compare with the existing method proposed in \cite{kurka2020deepjscc}, we comply with the loss function and the metric used in \cite{kurka2020deepjscc}, namely, the average mean squared error (MSE) and the average peak SNR (PSNR). The average MSE over N transmitted images is defined as follows:
 \begin{equation} 
\mathcal{L}=\frac{1}{N}\sum_{i=1}^Nd(\boldsymbol{x}^{(i)}, \boldsymbol{\hat{x}}^{(i)}),
\label{distortion_imgs}
\end{equation}
where $\boldsymbol{x}^{(i)}$ and $\boldsymbol{\hat{x}}^{(i)}$ represent the i-th original image and the corresponding reconstructed image, respectively, $d(\boldsymbol{x}^{(i)}, \boldsymbol{\hat{x}}^{(i)})$ is the MSE defined in Eq. \eqref{single_distortion}, and $N$ is the number of image samples. In practice, Eq. \eqref{distortion_imgs} insteads of Eq. \eqref{argmin} by assuming a given distribution of SNR and equally transmitting images in the dataset. In turn, the PSNR is defined as follows:
\begin{equation} 
\rm PSNR = 10log_{10}\frac{MAX^2}{MSE}(dB), 
\label{psnr}
\end{equation}
where MAX is the maximum possible value of the image pixels (e.g., MAX is 255 for the 8-bit color image). The PSNR is firstly calculated for each image and then averaged over all the tested images.

We use Tensorflow \cite{abadi2016tensorflow} and its high-level API Keras to implement the BDJSCC and ADJSCC\footnote{The following experiments demonstrate the ability of ADJSCC dealing with image source. With some appropriate modifications of the ADJSCC, the proposed mechanism can be applied to address other kinds of sources. However, it is out of our scope.} models. Consistently with the work \cite{kurka2020deepjscc}, Adam optimizer with a learning rate of $10^{-4}$ and the batch size with 128 is chosen to optimize the models. To measure the training efficiency, we let the training epochs to be euqal to 1280. Unless stated otherwise, CIFAR-10 \cite{krizhevsky2009learning} is used for training and evaluating the BDJSCC and ADJSCC models. The CIFAR-10 dataset consists of 60000 $32 \times 32 \times 3$ color images associated with 10 classes where each class has 6000 images. Note however we are not concerned with the class of each image because our goal is to reconstruct the original data from the received one with minimum distortion only. Training dataset and test dataset contain 50000 images and 10000 images, respectively. 

The BDJSCC method is trained at a specific SNR. In order to adapt to dynamic channel conditions, the ADJSCC method is trained under a uniform distribution within the SNR range [0, 20] dB. 
%We sample the channel SNR for each image in the dataset and then feed the batch of images and channel SNRs to train the ADJSCC network. 
Both the performance of the ADJSCC method and the BDJSCC method are evaluated under the specific SNR. Each image in the test dataset is transmitted 10 times to alleviate the effect caused by the randomness of the channel noise. All of our experiments are performed  on a Linux server with twelve octa-core Intel(R) Xeon(R) Silver 4110 CPUs and sixteen GTX 1080Ti GPU. Each experiment runs on six CPU cores and a GPU.

\subsection{ADJSCC Adaptability Experiments}
\label{subsection_4A}

 \begin{figure}[!htb]
\centering
\subfigure[]{
\label{awgn_compare_12}
\includegraphics[width=0.9\columnwidth]{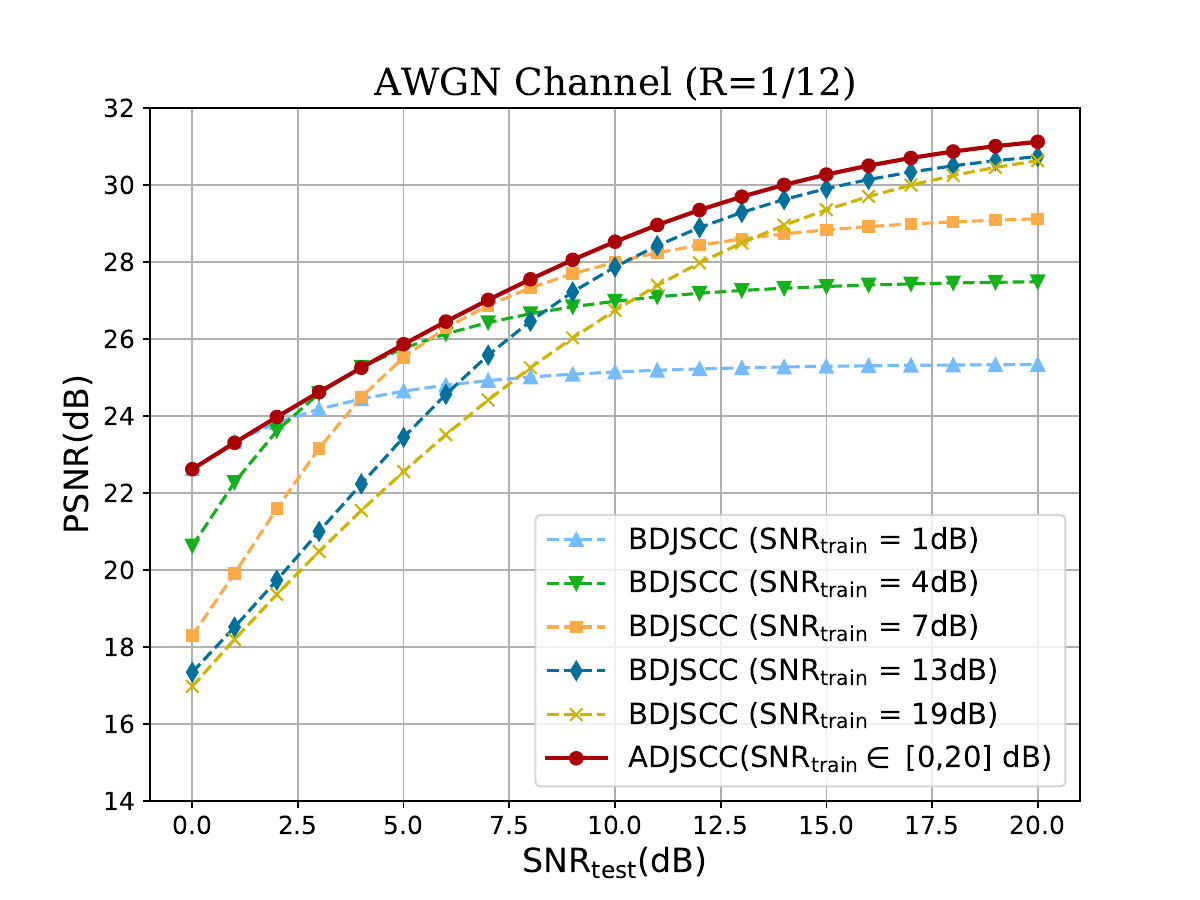}}
\subfigure[]{
\label{awgn_compare_6}
\includegraphics[width=0.9\columnwidth]{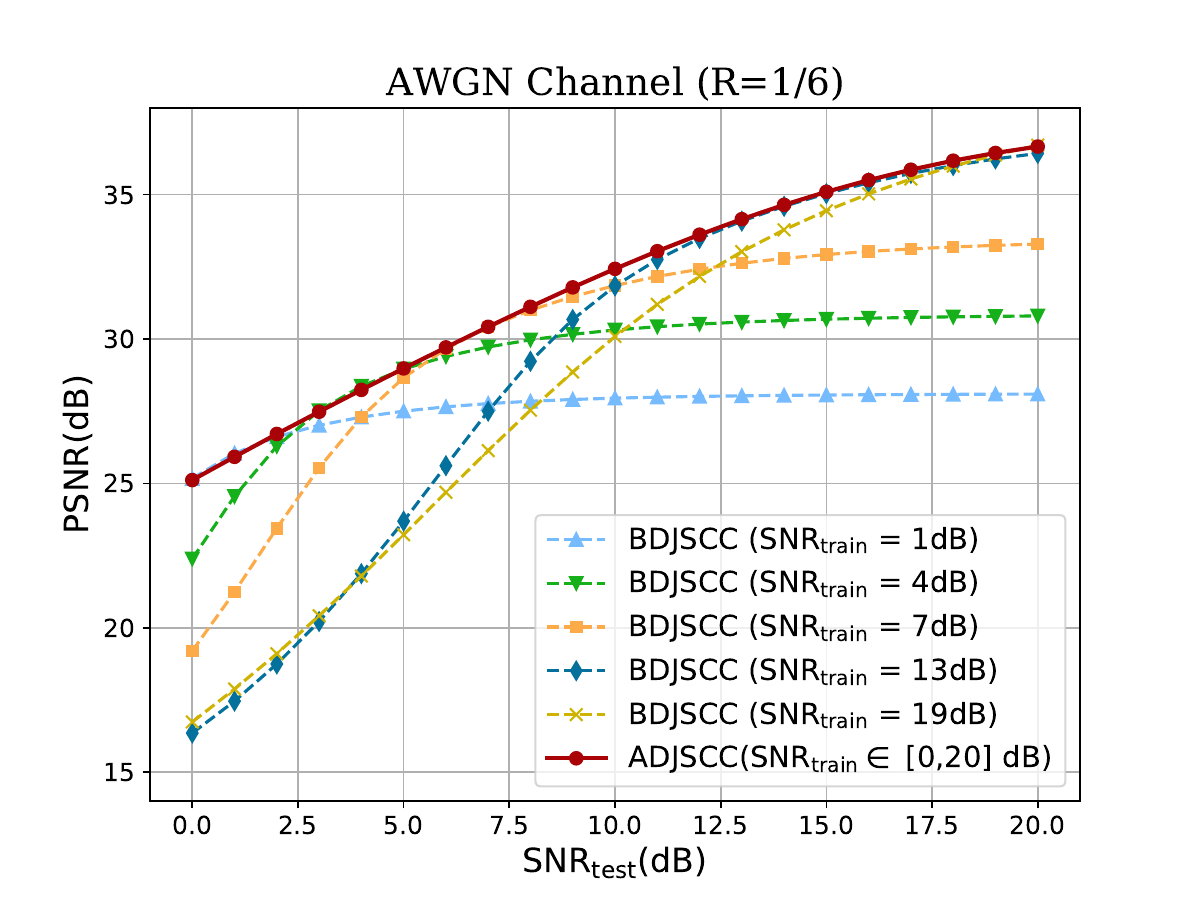}}

\caption{Performance of ADJSCC and BDJSCC on CIFAR-10 test images. (a) R =1/12 and (b) R=1/6. The curve of ADJSCC is trained under the uniform distribution of SNR from 0dB to 20dB. Each curve of BDJSCC is trained at a specific SNR.
}\label{Fig:awgn_compare}
\end{figure}

We first consider the performance of our proposed ADJSCC on the AWGN channel in Eq. \eqref{channel}. In the following experiment, the ADJSCC model is trained under the uniform distribution of $\rm SNR_{train}$ from 0 dB to 20 dB, and all of the BDJSCC models are trained at specific $\rm  SNR_{train}$ = 1dB, 4dB, 7dB, 13dB, 19dB, respectively, which are adopted in \cite{bourtsoulatze2019deep}. We however evaluate the performance of both ADJSCC and BDJSCC models at specific $\rm SNR_{test} \in$ [0,20] dB. 

Fig.~\ref{Fig:awgn_compare} compares the ADJSCC method with the BDJSCC method at bandwidth ratios $R$ = 1/12, 1/6, respectively. In Fig.~\ref{awgn_compare_12} with $R$ = 1/12, the performance of the ADJSCC model is better than the performance of any BJSCC model trained at the specific $\rm SNR_{train}$. With the increase of the $\rm SNR_{test}$, the ADJSCC model brings a gradually increased performance, outperforming the BDJSCC model ($\rm SNR_{train}$ = 1dB) by a margin of at most 6dB. With the decrease of the $\rm SNR_{test}$, the ADJSCC model still outperforms the BDJSCC model ($\rm SNR_{train}$ = 13dB or 19dB). It is also worth noting that the ADJSCC still outperforms the BDJSCC when $\rm SNR_{train}$ = $\rm SNR_{test}$\textemdash this is remarkable because ADJSCC is only trained under a certain SNR range. However, as $\rm SNR_{test}$ deviates from $\rm SNR_{train}$, the ADJSCC model tends to outperform considerably the BDJSCC one.

Fig.~\ref{awgn_compare_6} with $R$ = 1/6 reveals similar results to Fig.~\ref{awgn_compare_12}. However, with the increase of the bandwidth ratio, the performance gap between the ADJSCC model and the BDJSCC model ($\rm SNR_{train}$ = 13dB or 19dB) at high $\rm SNR_{test}$ regime almost disappears. Therefore, we conclude ADJSCC brings about higher performance gains than BDJSCC in the low bandwidth ratio regime. These results also suggest that using an ADJSCC is a much better strategy than using a collection of BDJSCCs where each BDJSCC is trained at a specific SNR level that are selected for transmission/reception depending on the actual channel conditions. Beyond the fact that such a strategy would lead to very complex transmitters and receivers, entailing considerable performance complexity, our results indicate that such a strategy does not lead to any performance gains over an ADJSCC strategy.

\begin{figure}[!htb]
\centering
\subfigure[]{
\label{en1_scale}
\includegraphics[width=0.8\columnwidth]{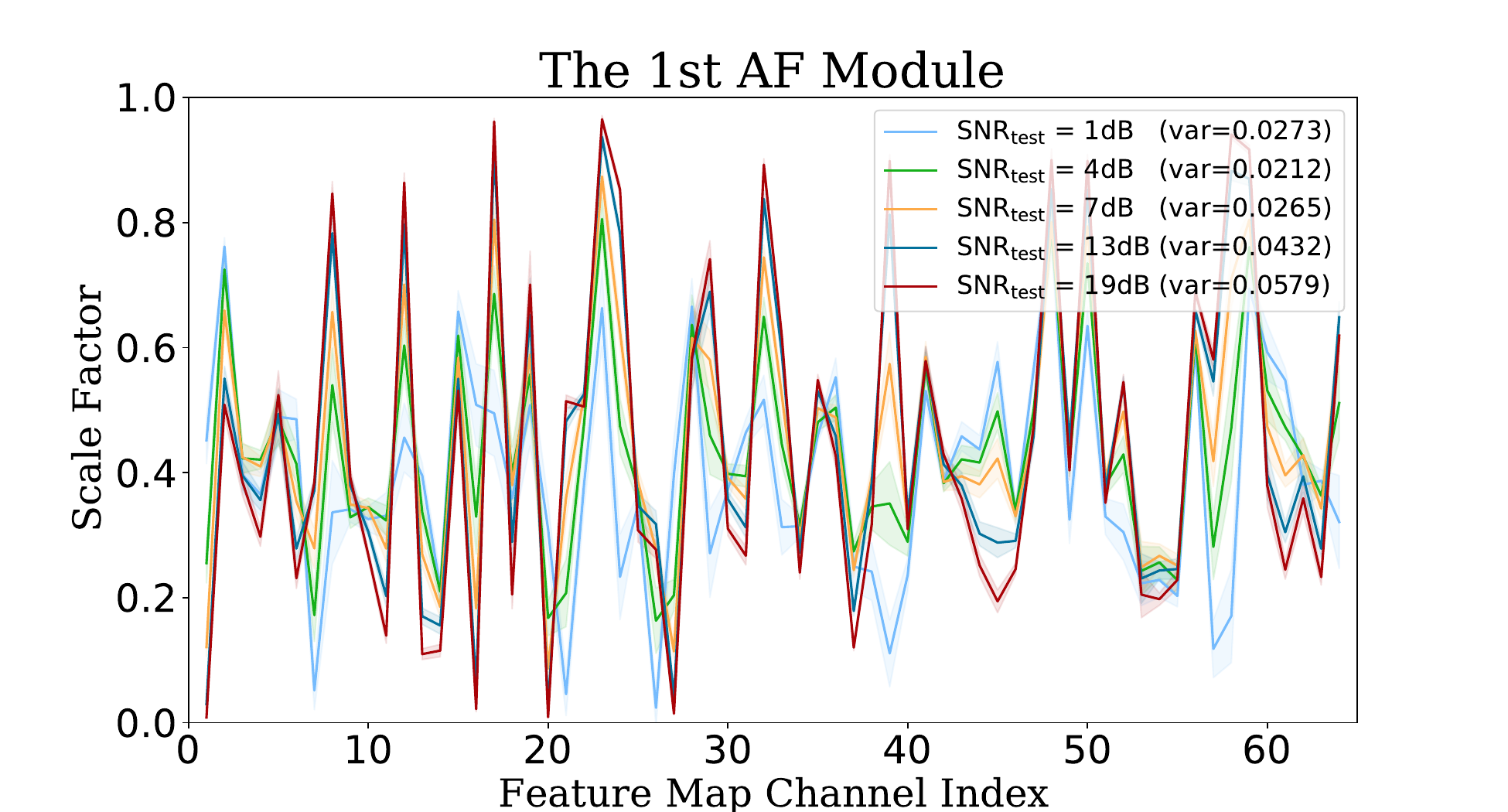}}
\subfigure[]{
\label{en2_scale}
\includegraphics[width=0.8\columnwidth]{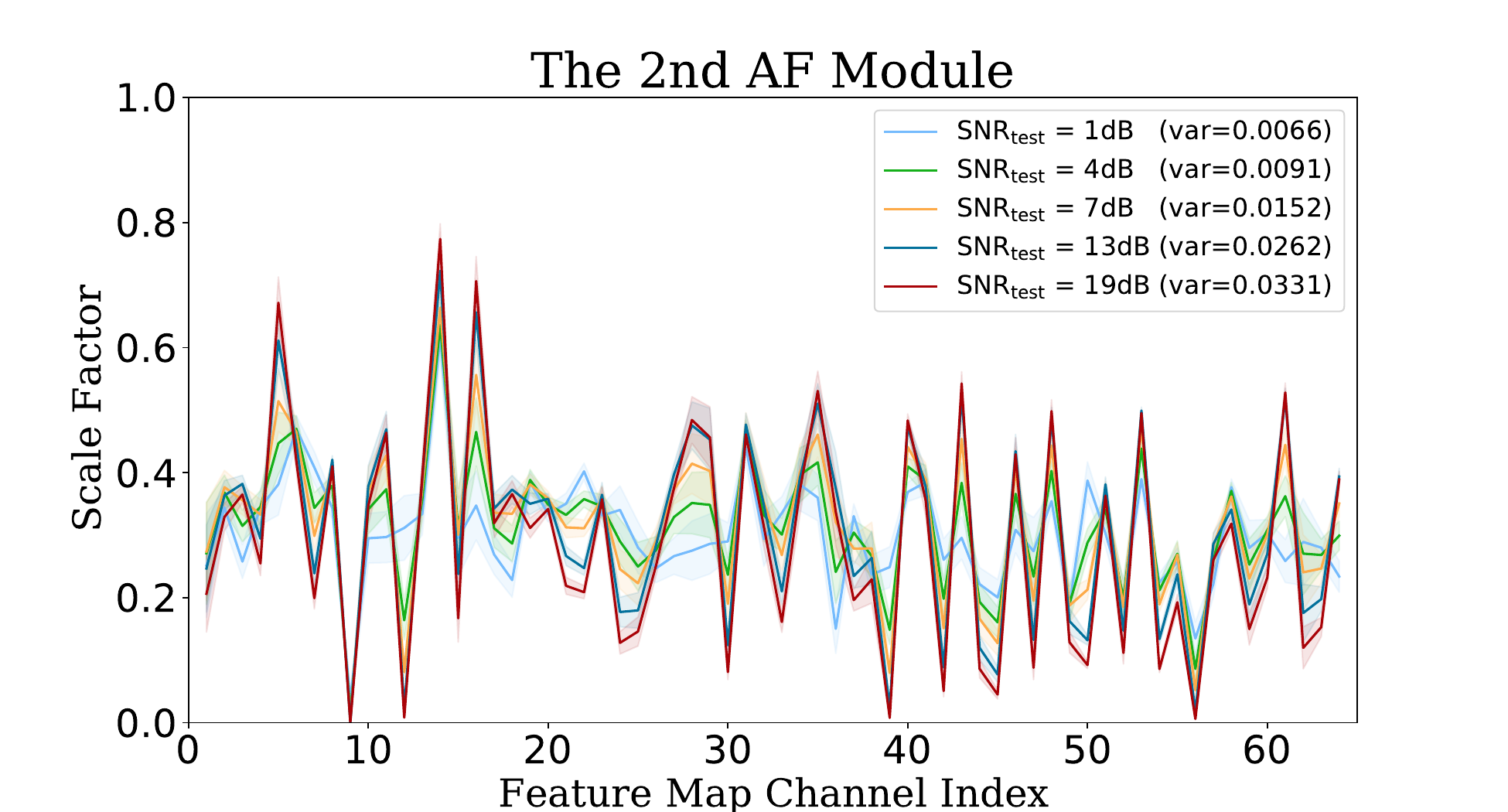}}
\subfigure[]{
\label{en3_scale}
\includegraphics[width=0.8\columnwidth]{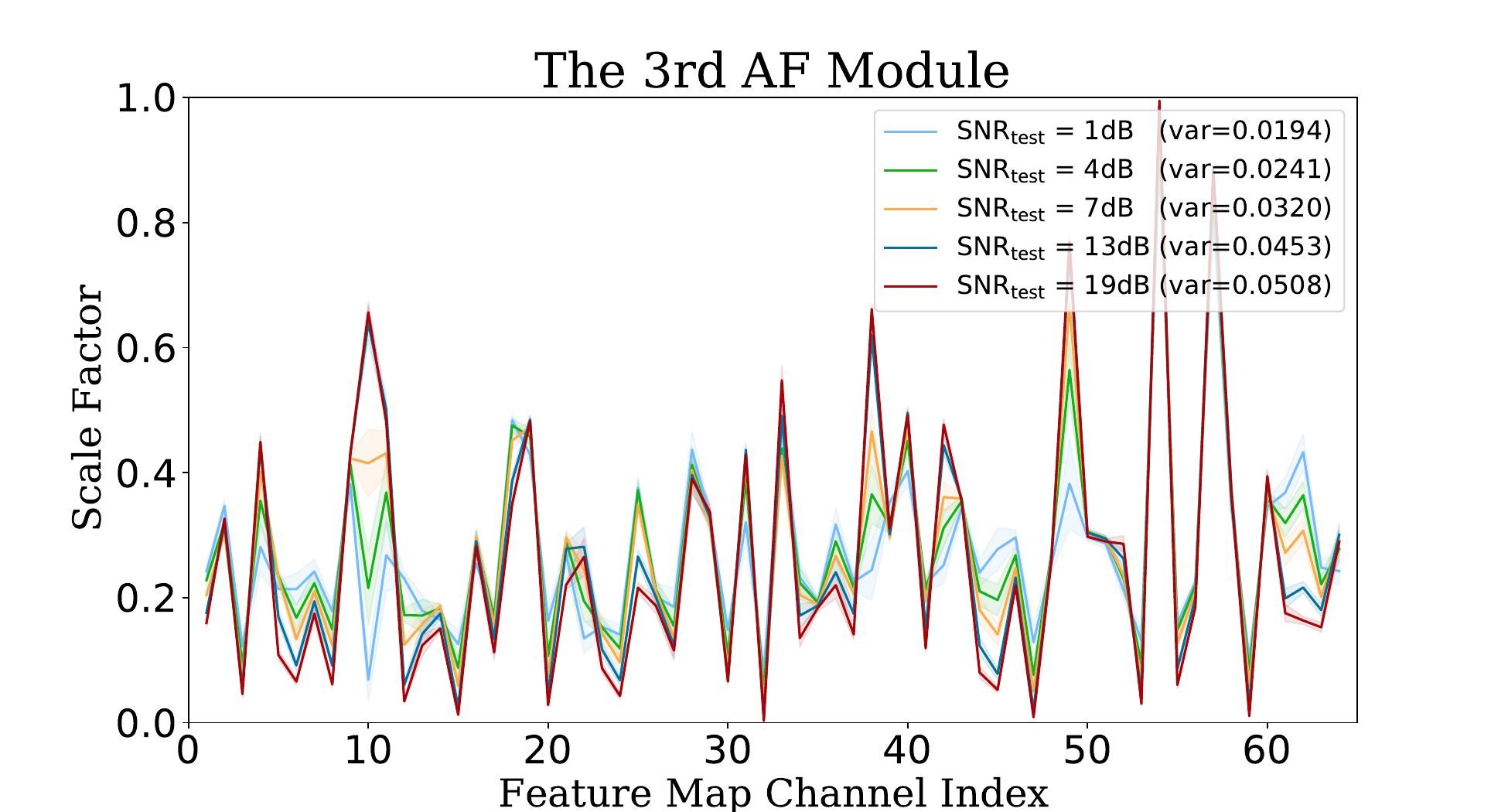}}
\subfigure[]{
\label{en4_scale}
\includegraphics[width=0.8\columnwidth]{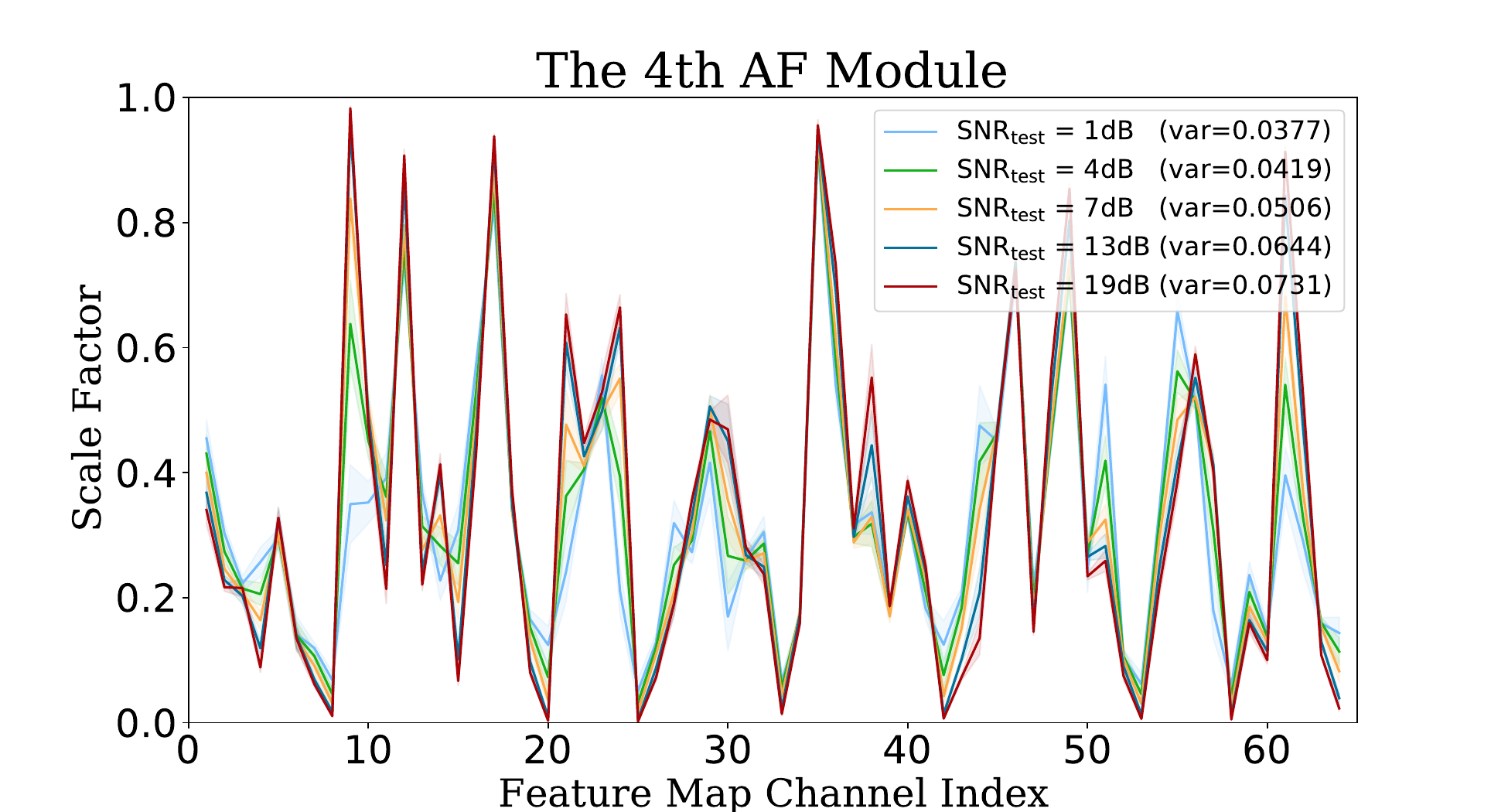}}
\caption{The scaling factors of the first 64 channels in the encoder of ADJSCC on AWGN channel (R=1/6). (a) the scaling factors of the 1st AF module, (b)the scaling factors of the 2nd AF module,  (c) the scaling factors of the 3rd AF module and  (d) the scaling factors of the 4th AF module. Each curve of the scaling factors is evaluated at a specific SNR on CIFAR-10 test dataset. The solid line represents the mean values of the scaling factors and the translucent area around the solid line with the same color represents the standard derivation of the scaling factors.The var represents the variance of the mean values of the scaling factors on specific channel SNR.}\label{Fig:scale_factor}
\end{figure}

\begin{figure*}[!htb]
\centering
\subfigure[]{
\label{en1_origin}
\includegraphics[width=0.5\columnwidth]{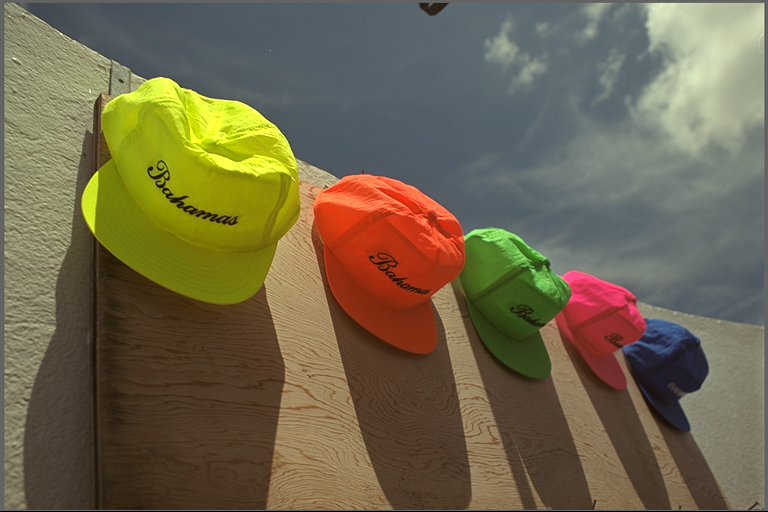}}
\subfigure[]{
\label{en1_snrdb1}
\includegraphics[width=0.6\columnwidth]{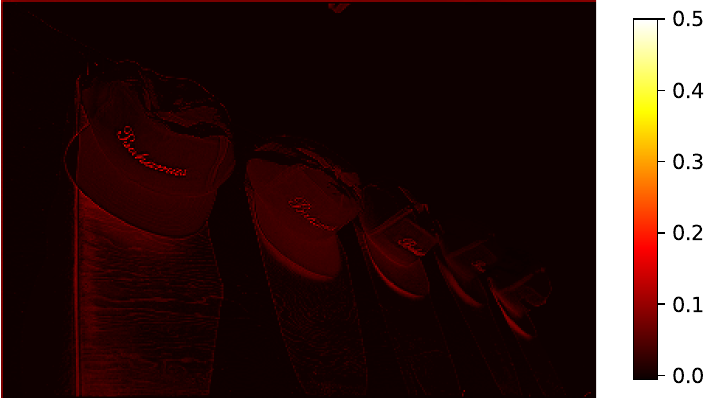}}
\subfigure[]{
\label{en1_snrdb19}
\includegraphics[width=0.6\columnwidth]{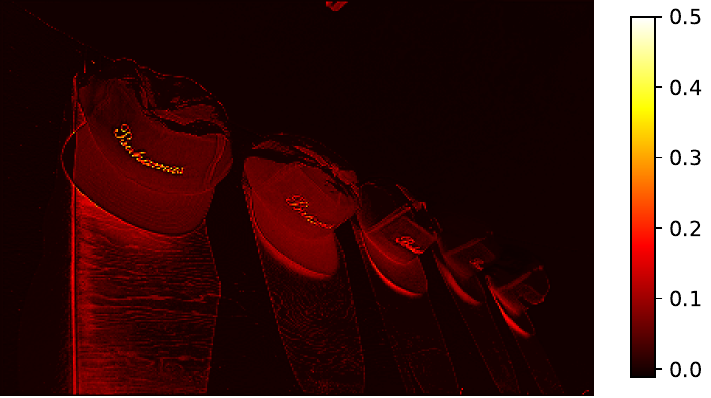}}
\caption{The original image and the heatmaps of the 23rd recalibrated feature in the first AM module of the encoder. (a) original image, (b) the heatmap at $\rm SNR_{test}$ = 1dB, (c) the heatmap at $\rm SNR_{test}$ = 19dB.}\label{Fig:en1_compare)}
\end{figure*}

The aforementioned performance evaluation of the ADJSCC model demonstrates  that the ADJSCC model can accommodate for a range of SNR by using attention mechanism. We would like to understand how the AF module affects the features in practice. To provide a thoughtful understanding of the AF module, we study the scaling factors generated by the AF module for the features. Specifically, we transmit the test dataset of CIFAR-10 at specific $\rm SNR_{test}$ and then compute the mean and standard deviation of the scaling factors for each AF module. We plot the mean and the standard deviation of the scaling factors in Fig.~\ref{Fig:scale_factor}. The solid line represents the mean values of the scaling factors and the translucent area around the solid line with the same color represents the standard deviation of the scaling factors. In order to show the scaling factors for each AF Model more clearly, we select the scaling factors of the first 64 channels in the encoder. From top to bottom, the rows correspond to the order from the first AF module to the forth AF module in the encoder. Two observations are inferred from Fig.~\ref{Fig:scale_factor}. First, the scaling factors across the channels fluctuate more drastically with the increase of the $\rm SNR_{test}$. This suggests that the scaling factors are more sensitive to high $\rm SNR_{test}$ than low $\rm SNR_{test}$, which coincides with the intuition. When the $\rm SNR_{test}$ is low, the channel noise is severe for each of the features. When the $\rm SNR_{test}$ is high, some of the features have a better contribution to the performance than others. The scaling factors of good features should be increased to improve the performance and the scaling factors of bad features should be decreased to avoid resource occupation.
Second, the difference between the curves at different $\rm SNR_{test}$ gets smaller as the AF module gets deeper. This observation shows that the channel noise has a more significant impact on low-level features than high-level features. Low-level features concentrate on the pixel relationship of an image. In contrast, high-level features concentrate more on the semantic representation implied in an image than the pixel relationship. Compared with the low-level features, high-level features are more robust to the channel noise. Therefore in the fourth AF module shown in Fig.~\ref{en4_scale}, the scaling factors at different $\rm SNR_{test}$ are similar. 

We also visualize the 23rd  feature of the first AF module in the encoder at $\rm SNR_{test}$ = 1dB shown in Fig.~\ref{en1_snrdb1} and $\rm SNR_{test}$ = 19dB shown in Fig.~\ref{en1_snrdb19}. The 32$\times$32 images of CIFAR-10 dataset are too small to be recognized by human eyes. Thus we use a 512$\times$768 image from Kodak dataset instead of the image from CIFAR-10 dataset. The detail of Kodak will be introduced in \ref{subsection_4C}.
Fig.~\ref{en1_origin} shows the original image. The comparison of Fig.~\ref{en1_snrdb1} and Fig.~\ref{en1_snrdb19} shows that when the channel exhibits high SNR, the information about caps should be enhanced to contribute more for the quality of the reconstructed image. It seems reasonable that the detailed information about the caps is easier to be disturbed as the channel gets worse. So when the channel exhibits high SNR, the detailed information about the caps should be decreased to save the transmit power for more robust information.  
%为了更清楚的展示通道的作用，我们选择前64个通道来做分析，观察图，得出2个结论，方差随着信噪比的变大而逐渐变小，意味着当SNR变大时，图像的背景信息起的作用越来越小，主要作用是由SNR决定的。当SNR变大时，缩放系数的波动程度越来越大，说明在缩放系数大的channel上对应的特征图对噪声是敏感的，而缩放系数始终在0.5左右的特征图，对噪声是不敏感的。

\subsection{
%Performance of the ADJSCC on Burst Channel
ADJSCC Robustness Experiments}
\begin{figure}[!htb]
\centering
\includegraphics[width=0.9\columnwidth]{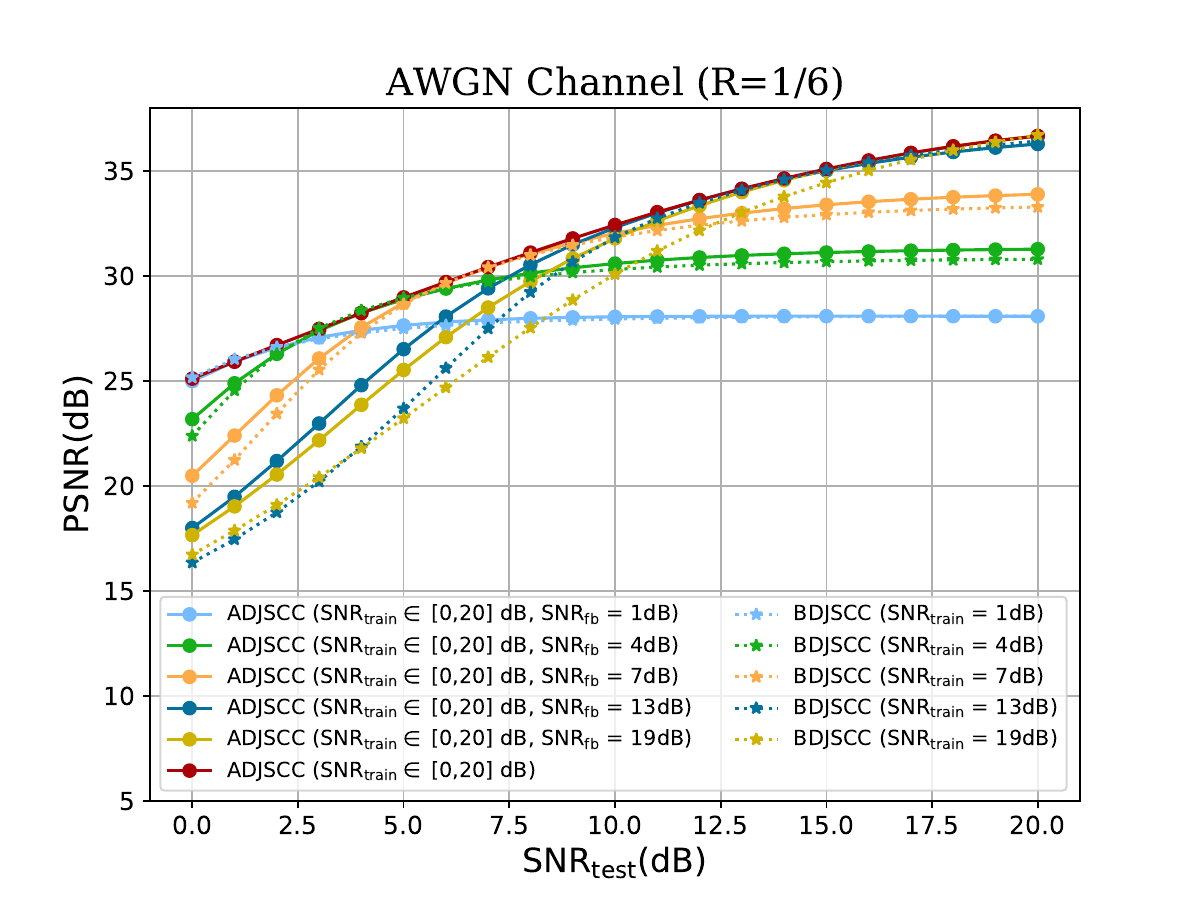}
\caption{Channel mismatch performance of ADJSCC and BDJSCC on CIFAR-10 test images.}
\label{c16_awgn_cifar10_mismatch}
\end{figure}

We now study the robustness of the proposed AJSCC scheme in the presence of channel mismatch. Fig.~\ref{c16_awgn_cifar10_mismatch} shows the mismatch performance of the ADJSCC model and the BDJSCC model under the AWGN channel with $R$ = 1/6. Compared with the case without channel mismatch, the ADJSCC models have some performance loss in channel mismatch conditions. The ADJSCC models still outperforms the BDJSCC models in the case of channel mismatch. The ADJSCC model shows better robustness especially in high $\rm SNR_{fb}$ regimes. The gain of the ADJSCC against the BDJSCC is higher in the case when the feedback SNR is large.

\subsection{ADJSCC Versatility}
\label{subsection_4C}

\begin{figure}[!htb]
\centering
\includegraphics[width=0.9\columnwidth]{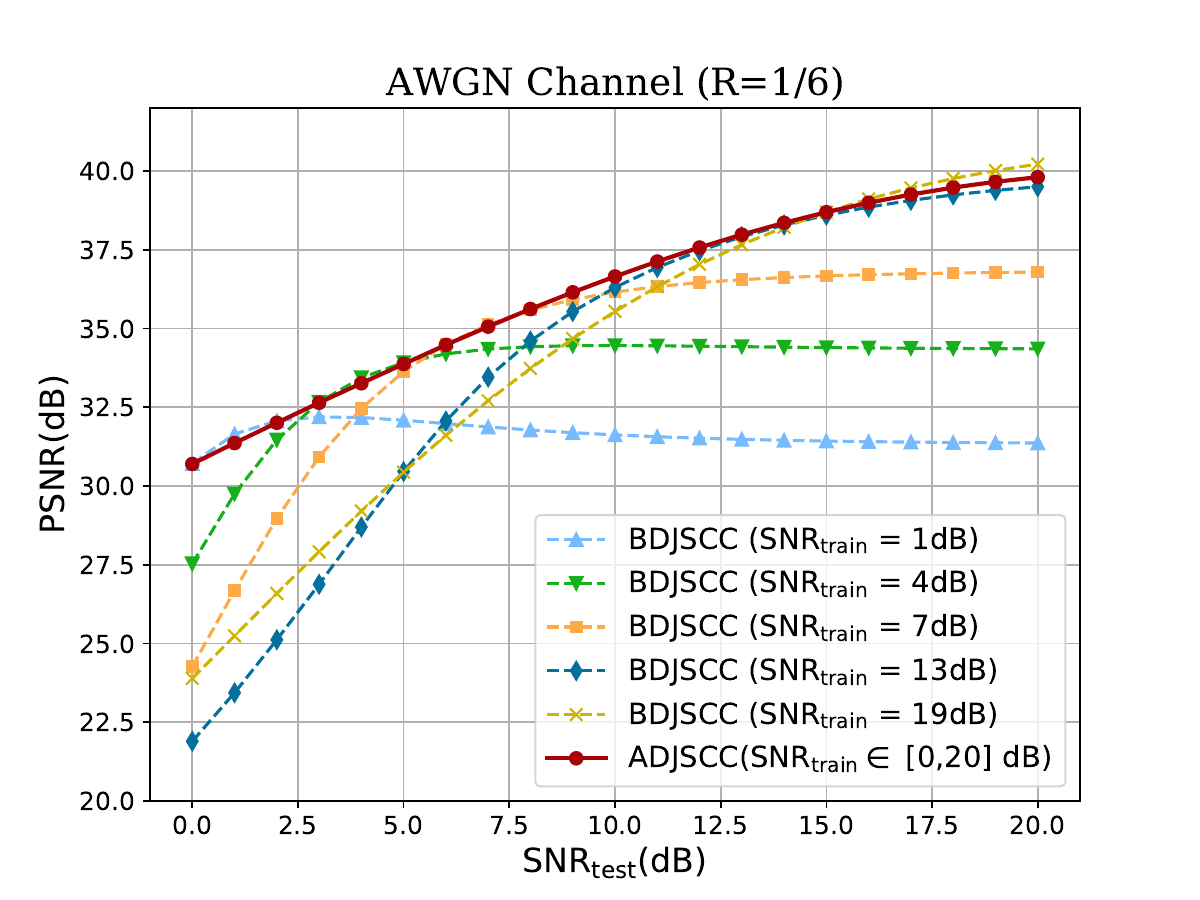}
\caption{Performance of ADJSCC and BDJSCC on Kodak dataset. ADJSCC and BDJSCC are trained on Imagenet dataset for bandwidth ratio R=1/6. The curve of ADJSCC is trained under the uniform distribution of SNR from 0dB to 20dB. Each curve of BDJSCC is trained at a specific SNR.}
\label{Fig:compare_imagenet}
\end{figure}

\begin{figure*}[!htb]
\centering
\includegraphics[width=2\columnwidth]{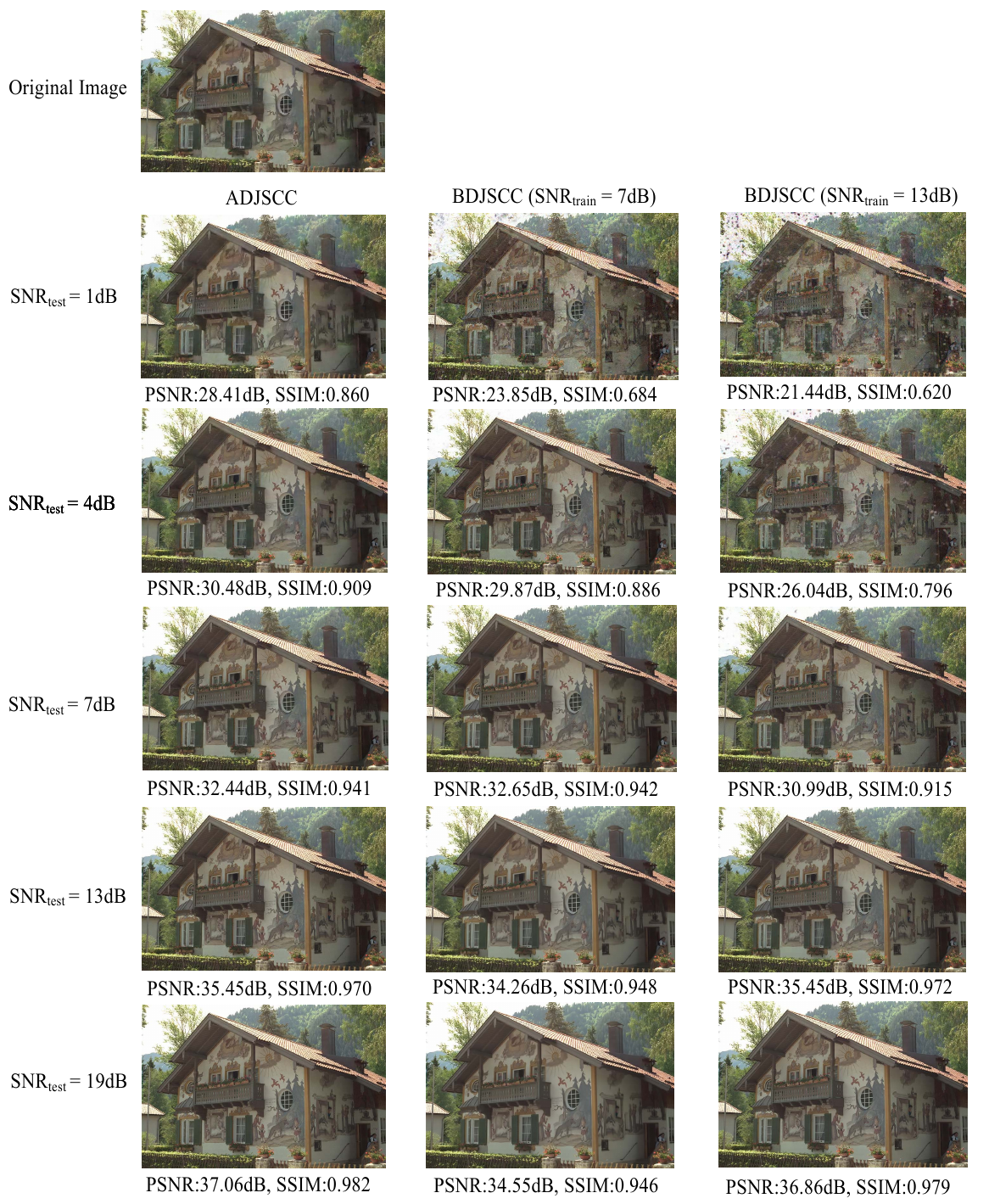}
\caption{Visual comparison between the ADJSCC model and the BDJSCC models ($\rm SNR_{train}=7dB, 13dB$) for the sample image of the Kodak dataset. We also used a perceptual metric\textemdash structural similarity index (SSIM)\textemdash to evaluate image quality. }\label{Fig:visual_present}
\end{figure*}

The effectiveness of ADJSCC has been demonstrated on CIFAR-10 dataset in Section \ref{subsection_4A}. However, CIFAR-10 images exhibit low-resolution. Therefore, we also now test the performance of our proposed approach on higher resolution images as in \cite{bourtsoulatze2019deep} and \cite{kurka2020deepjscc}. It is worth noting that the duration of coded symbols transmitted by the transmitter need to be smaller than the coherence time to promise the constant SNR during the transmission for real wireless scenarios. For fading channels (e.g., Rayleigh channel) which suit for the large PHY block case, one can train the deep neural network with properly prepared channel.

We train ADJSCC on ImageNet dataset\footnote{Until now, ImageNet dataset contains more than 14 million images with more than 21 thousand classes.} \cite{russakovsky2015imagenet} and evaluate ADJSCC on Kodak dataset. The ImageNet dataset contains various images of various sizes. We choose images with size larger than 128$\times$128 and then randomly crop them to a size of  128$\times$128 to generate the training dataset (About 5.8 million images satisfy this condition). Kodak dataset consists of 24 768$\times$512 images. Two epochs with batch size of 16 and learning rate of 0.0001 is enough to make the ADJSCC model converge. The model is saved every 200 training batches. To average the random of the channel, each images in Kodak is transmitted 100 times to evaluate the ADJSCC performance. Note that it seems not rational to train our model on a dataset and evaluate our model on another dataset. However, training on a sufficiently complex dataset (e.g., ImageNet dataset) allows to perform well on other datasets (e.g., Kodak dataset). In addition, the size of evaluation dataset (768$\times$512 in Kodak dataset) can be different from that of training dataset (128$\times$128 in Kodak dataset) owes to the full convolutional architecture adopted in ADJSCC. Different from classical CNN networks\footnote{Classical CNN networks (e.g., Alexnet) use the full connected layer after convolution layers to get fixed length feature vector for classification.}, this full convolutional architecture only uses the convolutional layers to extract image features and uses transposed convolutional layers to restore the image, which allows different size of the input image as long as it meets the requirements of stride (the image size $n$ must be a multiple of 4).

%ADJSCC is a full convolutional architecture, which is different from classical CNN networks (e.g., Alexnet) that use the full connected layer after the convolution layer to get fixed length feature vector for classification. The classical CNN networks require that the training dataset and the test dataset have the same image size. the full convolutional architecture uses the convolutional layer instead of the full connected layer in classical CNN networks, which can accept different size of the input image as long as it meets the requirements of stride (e.g., the image size $n$ must be a multiple of 4) and use the transposed convolutional layer to restore to the same size of the input image.  

Fig.~\ref{Fig:compare_imagenet} shows the comparison of the ADJSCC model and the BDJSCC models on Kodak dataset. If we consider the BDJSCC models as a whole, the performance of the ADJSCC model approaches that of the ensemble of  BJSCC models when $\rm SNR_{test} \leq 17dB$ with negligible difference. Moreover, the performance of the ADJSCC model is 0.3dB lower than that of the ensemble of BJSCC models when $\rm SNR_{test} > 17dB$. Note however that the regime $\rm SNR_{test} > 17dB$ is associated with images with $\rm PSNR > 35dB$ whose quality is virtually impossible to distinguish with human eyes\footnote{Note also that using an ensemble of BJSCC models in lieu of a single ADJSCC model is problematic because\textemdash as discussed earlier\textemdash it requires storing the different networks at transmitter and receiver, as well as switching between one or another depending on SNR conditions. This has considerably storage demands preventing the use of an ensemble of BJSCC models in various applications.}. 

 We present a visual comparison between the ADJSCC model and the BDJSCC models for the sample image of the Kodak dataset in Fig.~\ref{Fig:visual_present}. Note that, even if the $\rm SNR_{test}$ is the same, the reconstructed image is different at each time due to the randomness of the noise. 
%Moreover, the versatility of ADJSCC is only demonstrates the effectiveness of the ADJSCC method training at a general dataset and test at a special dataset without strict image size constraints.
%The training dataset is important for the performance of ADJSCC. 

\begin{figure}[t]
\centering
\includegraphics[width=0.9\columnwidth]{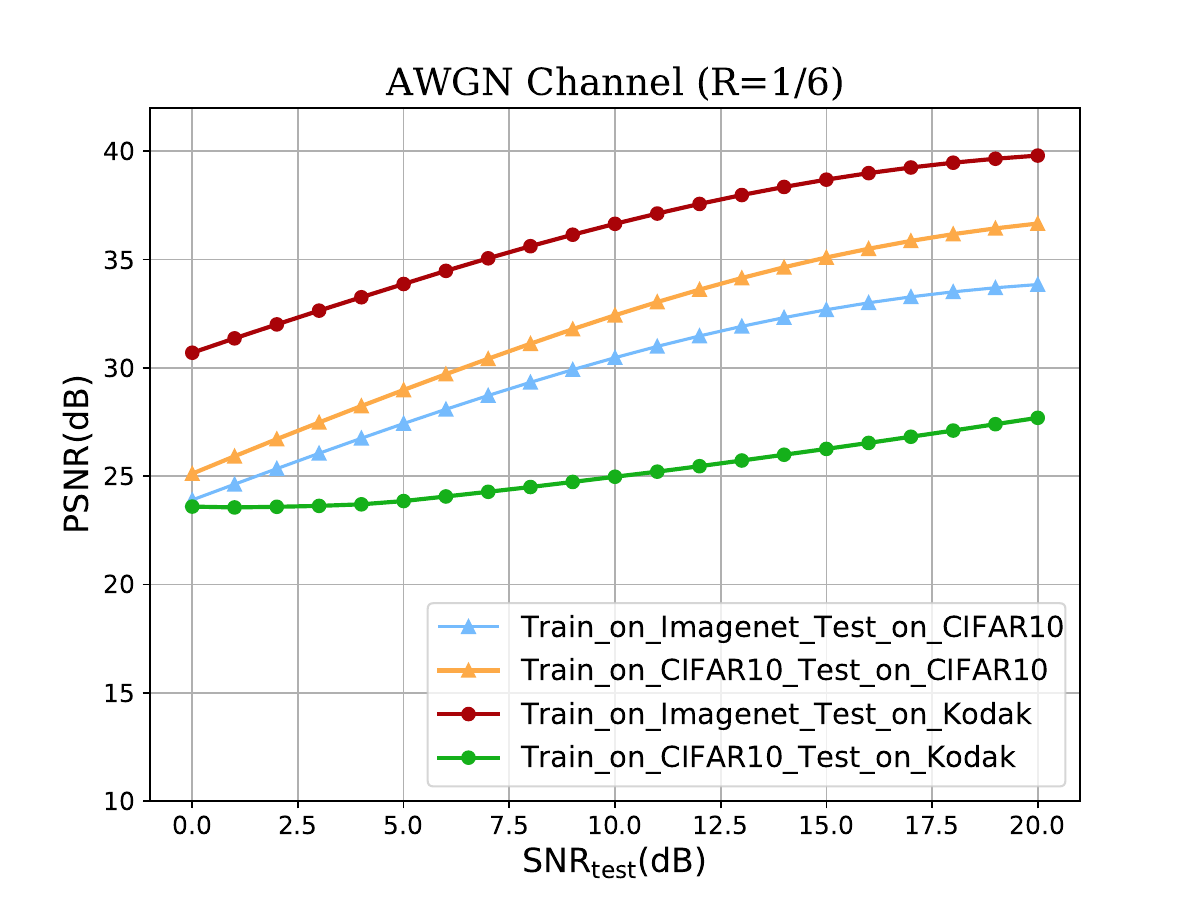}
\caption{Performance of the ADJSCC methods training on Imagenet or CIFAR-10 and testing on Kodak or CIFAR-10 for bandwidth ratio R=1/6. The ADJSCC methods are trained under the uniform distribution of SNR from 0dB to 20dB.}
\label{Fig:compare_cifar10_kodak}
\end{figure}

To further demonstrate the rationale of this experiment, we compare the performance of ADJSCC models trained on Imagenet or CIFAR-10 and tested on Kodak or CIFAR-10 in Fig.~\ref{Fig:compare_cifar10_kodak}. The performance of the ADJSCC model trained on CIFAR-10 and tested on CIFAR-10 is better than the performance of the ADJSCC model trained on Imagenet and tested on CIFAR-10, which is consistent with the traditional requirement of DL that the test dataset should have the similar distribution with the training dataset to get good performance. Compared with the ADJSCC model trained on CIFAR-10 and tested on CIFAR-10, the ADJSCC model trained on Imagenet and tested on CIFAR-10 only has 1dB to 3dB performance loss in the range of SNR from 0dB to 20dB. However, the performance of the ADJSCC model trained on CIFAR-10 and tested on Kodak is much worse than the performance of the ADJSCC model trained on Imagenet and tested on Kodak. The performance loss caused by the image size mismatch is limited when the ADJSCC model is trained on a large number of higher resolution images (e.g., Imagenet), while there is a huge performance loss caused by the image size mismatch when the ADJSCC model is trained on a lower resolution images (e.g., CIFAR-10).  
 %The comparison of Fig.~\ref{Fig:compare_imagenet} reveals that our proposed ADJSCC method is also applicable to large dataset. 

\begin{figure}[t]
\centering
\includegraphics[width=0.9\columnwidth]{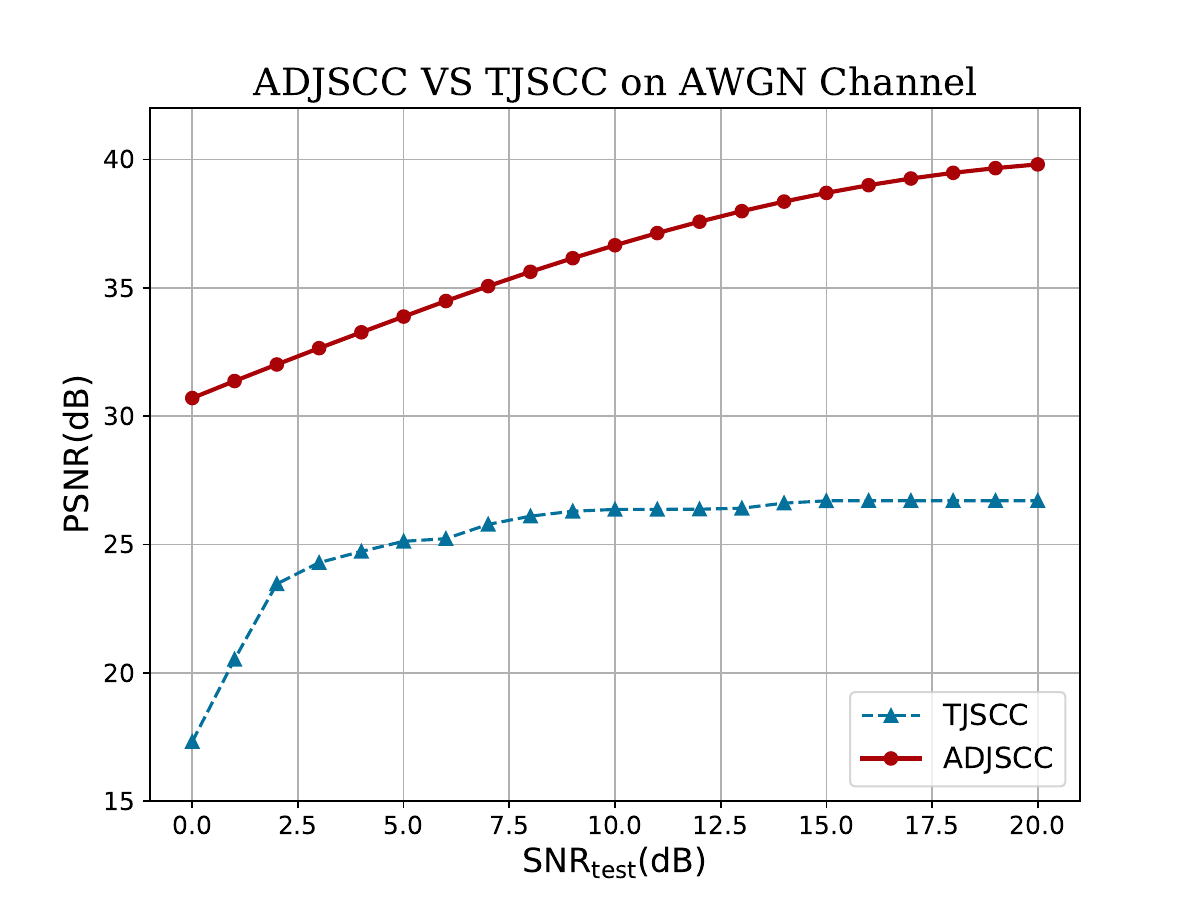}
\caption{Performance of ADJSCC and TJSCC on Kodak dataset. ADJSCC is trained on ImageNet dataset and  with respect to the uniform distribution of SNR from 0dB to 20dB for bandwidth ratio R=1/6. The bandwidth ratio of the TJSCC $R_t =2R = 1/3.$}
\label{Fig:tjscc_vs_adjscc}
\end{figure}

We finally compare our proposed method with the traditional JSCC (TJSCC) method proposed in \cite{sprljan2005fast}, which is an unequal error protection scheme for lossy channels and fix packet size transmission. TJSCC uses Set Partitioning in Hierarchical Trees (SPIHT) coder as a source encoder to encode the original image to the embedded source bits. Cyclic redundancy check (CRC) bits are appended to the source bits to enhance the protection. Rate-compatible punctured convolutional (RCPC) codes are used as channel encoder and combined with fast unequal protection scheme to encode the embedded source bits to channel bits depending on the bandwidth ratio $R_t$, the rate-distortion of the original image and the channel status. BPSK modulation is assumed in TJSCC. Other settings are consistent with the setting in \cite{sprljan2005fast}. Because TJSCC is only fit for gray images, we separate the color images of Kodak into three channels. Each of the three channels is processed by TJSCC. The channel output symbols of ADJSCC are complex valued. However, the channel output symbols of TJSCC are real valued. For the sake of fairness, the bandwidth ratio of TJSCC is $R_t$ = 2$R$, where $R$ is the bandwidth ratio of ADJSCC. Fig.~\ref{Fig:tjscc_vs_adjscc} shows the comparison of ADJSCC  (trained on Imagenet dataset)  and TJSCC on Kodak dataset. 
The performance improvement of TJSCC is fast from $\rm SNR_{test}=0dB$ to $\rm SNR_{test}=2dB$ and becomes slow from $\rm SNR_{test}=3dB$ to $\rm SNR_{test}=15dB$. When $\rm SNR_{test} > 15dB$, the performance of TJSCC is saturated. Compared with the TJSCC method, the ADJSCC method brings a large performance improvement. The minimum performance improvement is around 7dB at $\rm SNR_{test}=3dB$ and the maximum performance improvement is around 13 dB at $\rm SNR_{test}=20dB$. Besides the performance advantage, ADJSCC does not need to calculate the rate-distortion of the original image, resulting in a huge computational demand at the transmitter as shown in \cite{sprljan2005fast}.

\section{Storage Overhead and Computational Complexity}
% Fig.~\ref{awgn_compare_6} reveals that the performance of the proposed ADJSCC can achieve or surpass the performance the whole BJSCC under the condition that only a fraction of BJSCC's storage is required. The relationship between the storage and the performance of ADJSCC and BDJSCC with $R=1/6$ are listed in Table~\ref{Table:one_block}. %解释容量大小和PSNR性能之间的关系。
We finally calculate the storage overhead required by the BDJSCC model and the ADJSCC model with $R$ = 1/6. The storage overhead of both ADJSCC and BDJSCC are independent of the image size. There are 10,690,351 total parameters in the BDJSCC model. The type of each parameter is float32, which requires 4 bytes. Hence the storage of the BDJSCC model is 10,690,351$\times$4$\approx$40.78 MB. The ADJSCC model has more parameters than the BDJSCC model because of the existing AF modules. The amount of the total parameters of the ADJSCC model is 10,758,191. The storage of the ADJSCC model is almost 41.04 MB. The ADJSCC model has 0.6\% more parameters than the BDJSCC model, which needs a little additional storage. 

However, we have already discussed that in practice one would desire to use an ensemble of BDJSCC models, each trained at a particular $\rm SNR_{train}$ and tested at a $\rm SNR_{test}$ = $\rm SNR_{train}$. Let us denote by BDJSCC-1 as a strategy involving using one BJSCC model trained at $\rm SNR_{train}$ = 10dB that is evaluated by the CIFAR-10 dataset with the assumed $\rm SNR_{test}$. Let us in turn denote by BDJSCC-2 as a strategy involving using two BJSCC models trained at $\rm SNR_{train}$ = 5dB, 15dB that are evaluated by the CIFAR-10 dataset with the assumed $\rm SNR_{test}$. In particular, this strategy involves selecting between one BJSCC model trained at $\rm SNR_{train}$ = 5dB and the other trained at $\rm SNR_{train}$ = 15dB depending on the difference between $\rm SNR_{train}$ and $\rm SNR_{test}$.
Finally, we let BDJSCC-5 denote a strategy involving using five models trained at $\rm SNR_{train}$ = 2dB, 6dB, 10dB, 14dB, 18dB, respectively, and BDJSCC-10 the strategy involving using ten models trained at $\rm SNR_{train}$ = 1dB, 3dB, 5dB, 7dB, 9dB, 11dB, 13dB, 15dB, 17dB, 19dB, respectively. Table \ref{Table:evaluation} compares the performance/storage demands associated with each of these strategies. Note that BDJSCC-1 has a slightly lower storage overhead than ADJSCC, but it considerably underperforms ADJSCC by 5.3 dB. The BDJSCC-10 has similar performance with the ADJSCC. However, the storage overhead is 10 times that of the ADJSCC. 

\begin{table}[!tb] 
\renewcommand{\arraystretch}{1.3} 
\caption{Evaluation of the ADJSCC Model and BDJSCC Model Strategies} 
\label{Table:evaluation}
\centering
\begin{tabular}{c|c|c}
\hline
\bfseries Strategy Name & \bfseries Storage & \bfseries PSNR\\ \hline
ADJSCC & 41.04MB & \bfseries 29.831dB\\ \hline
BDJSCC-1 & \bfseries 40.78 MB & 24.474dB\\ \hline
BDJSCC-2 & 81.56MB & 28.495dB\\ \hline
BDJSCC-5 & 203.9MB & 29.694dB\\ \hline
BDJSCC-10 & 407.8MB & 29.826dB \\ \hline
\end{tabular}
\end{table}

Then we evaluate the computational complexity of the ADJSCC model and the BDJSCC model on the Linux Server we mentioned in Section \ref{s4}. With a training mini-batch of 128 images from CIFAR-10 training dataset, the mean training time of the ADJSCC model for a batch is almost 114 ms whereas the mean training time of the BDJSCC model for a batch is almost 110 ms. The inference time for the ADJSCC model takes 53 ms, compared to 49 ms for the BDJSCC model on CIFAR-10 dataset images. In summary, the training time of the ADJSCC model is 3.6\% higher than that of the BDJSCC model and the inference time of the ADJSCC model is 8.1\% higher than that of the BDJSCC model. However, note again that one would in practice have to use multiple BDJSCC models in order to achieve a performance comparable to the ADJSCC model, as discussed earlier, so, all in all, our proposed approach exhibits a much better computational/storage complexity that BDJSCC.

\section{Conclusion}
In this work, we have proposed a novel ADJSCC method based on attention mechanisms that can adapt automatically to various channel conditions. It exhibits better performance, computational complexity, and storage complexity than existing approaches, making it an ideal candidate for JSCC in practical wireless communications scenarios.

%We first introduce the channel-wise soft attention, a method that can keep the differentiability of the neural network to execute the backpropagation algorithm. Then 
We have proposed the ADJSCC method to be built upon two types of modules: the FL modules and the AF modules. The FL module is a general module that can utilize the existing module designed in the existing DJSCC work. The AF module takes the context information to generate the scaling factors by using the channel-wise soft attention and then recalibrates the channel-wise features. The motivation of our ADJSCC method originates from the resource assignment strategy in the traditional concatenated source channel coders. 

We have evaluated the proposed ADJSCC method on AWGN channel, channel mismatch and large dataset to demonstrate the adaptability, the robustness and the versatility of the ADJSCC method. Lastly, we have compared the storage overhead and computational complexity of the ADJSCC method with that of the BDJSCC method. To achieve the same performance (PSNR), ADJSCC only needs  10.06\% of the storage required by BDJSCC-10 and 10.36\% of the training time required by BDJSCC-10.

In future work, a potential direction is to extend the proposed method for high-definition images and real wireless channels. This will play an important part in promoting the DL based JSCC technology for practical wireless communication systems.

% or


%\appendices
%\section{Image Reconstructions from different schemes}
%In order to provide a visual comparison of our algorithm versus traditional digital separation-based schemes, we present sample images from the Kodak dataset, produced on a setup on AWGN channel, SNR=1dB and bandwidth compression R = 1/6. We compare DeepJSCC-f (noiseless feedback) with separation-based scheme using 1/3 rate LDPC + 4QAM as channel code and JPEG, JPEG2000 and BPG as source codes. Results can be seen in Figures 17 and 18, where the PSNR, SSIM and MS-SSIM performances are computed.

%\section{}

%\section*{Acknowledgment}


\ifCLASSOPTIONcaptionsoff
  \newpage
\fi


\bibliographystyle{IEEEtran}
\bibliography{ref/IEEEabrv,ref/ref}


%\begin{comment}
\begin{IEEEbiography}[{\includegraphics[width=1in,height=1.25in,clip]{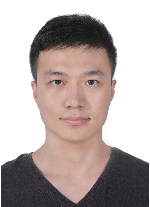}}]{Jialong Xu }
(Student Member, IEEE) received the B.E. and M.S. degrees from Engineering University of PAP in 2009 and 2012 respectively. He is currently pursuing the Ph.D. degree with the State Key Laboratory of Rail Traffic Control and Safety, Beijing Jiaotong University, Beijing, China. His research interests include deep learning, wireless coding and information theory.
\end{IEEEbiography}

\begin{IEEEbiography}[{\includegraphics[width=1in,height=1.25in,clip]{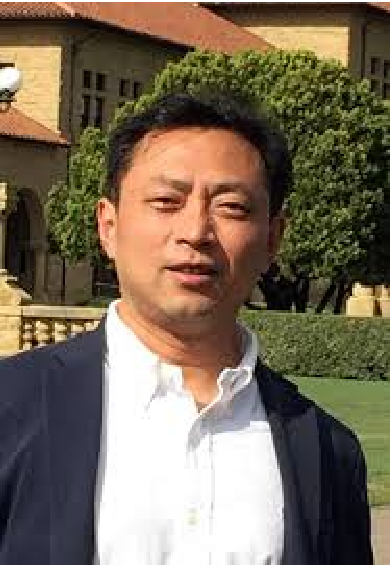}}]{Bo Ai}(Senior Member, IEEE) received the M.S. and Ph.D.degrees from Xidian University, Xian, China, in 2002 and 2004, respectively. He was with Tsinghua University, Beijing, China, where he was an Excellent Postdoctoral Research Fellow in 2007. He is currently a Professor and an Advisor of Ph.D.candidates with Beijing Jiaotong University, Beijing, where he is also the Deputy Director of the State Key Laboratory of Rail Traffic Control and Safety. He is also currently with the Engineering College, Armed Police Force, Xian. He has authored or coauthored six books and 270 scientific research papers, and holds 26 invention patents in his research areas. His interests include the research and applications of orthogonal frequency-division multiplexing techniques, high-power amplifier linearization techniques, radio propagation and channel modeling, global systems for mobile communications for railway systems, and long-term evolution for railway systems.

Dr. Ai is a Fellow of The Institution of Engineering and Technology. He was as a Co-chair or a Session/Track Chair for many international conferences such as the 9th International Heavy Haul Conference (2009); the 2011 IEEE International Conference on Intelligent Rail Transportation; HSRCom2011; the 2012 IEEE International Symposium on Consumer Electronics; the 2013 International Conference on Wireless, Mobile and Multimedia; IEEE Green HetNet 2013; and the IEEE 78th Vehicular Technology Conference (2014). He is an Associate Editor of IEEE TRANSACTIONS ON CONSUMER ELECTRONICS and an Editorial Committee Member of theWireless Personal Communications journal. He has received many awards such as the Qiushi Outstanding Youth Award by HongKong Qiushi Foundation, the New Century Talents by the Chinese Ministry of Education, the Zhan Tianyou Railway Science and Technology Award by the Chinese Ministry of Railways, and the Science and Technology New Star by the Beijing Municipal Science and Technology Commission.
\end{IEEEbiography}

\begin{IEEEbiography}[{\includegraphics[width=1in,height=1.25in,clip]{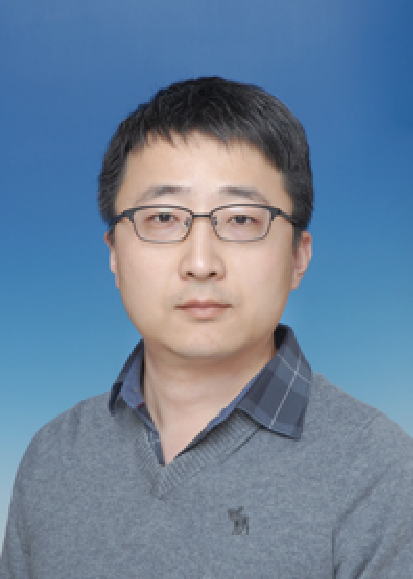}}]{Wei Chen}
(Senior Member, IEEE) received the B.Eng. and M.Eng. degrees in communications engineering from the Beijing University of Posts and Telecommunications, Beijing, China, in 2006 and 2009, respectively, and the Ph.D. degree in computer science from the University of Cambridge, Cambridge, U.K., in 2013. He was a Research Associate with the Computer Laboratory, University of Cambridge from 2013 to 2016. He is currently a Professor with Beijing Jiaotong University, Beijing. His current research interests include sparse representation, Bayesian inference, wireless communication systems and image processing. He was the recipient of the 2013 IET Wireless Sensor Systems Premium Award and the 2017 International Conference on Computer Vision (ICCV) Young Researcher Award.
\end{IEEEbiography}

\begin{IEEEbiography}[{\includegraphics[width=1in,height=1.25in,clip]{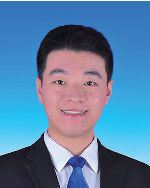}}]{Ang Yang}
received the B.S. and Ph.D. degree in electronic engineering from Beijing Institute of Technology (BIT), Beijing, China, in 2009 and 2015, respectively.
From 2015 to 2018, he worked as an algorithm engineer with Samsung Beijing Research Center. Since 2018, he has been a standard engineer with vivo Communication Research Institute. His research interests include massive multiple-input–multiple-output systems, millimeter wave and artificial intelligence.
\end{IEEEbiography}

\begin{IEEEbiography}[{\includegraphics[width=1in,height=1.25in,clip]{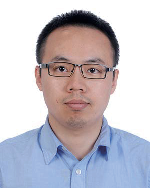}}]{Peng Sun}
received the B.S., M.S. and Ph.D. degree in electronic engineering from Beijing University of Posts and Telecommunications, Beijing, China, in 2006, 2009 and 2013 respectively. From 2012 to 2017 June, he worked as a PHY/MAC system engineer with Beijing Xinwei Telecommunication Company. Since 2017 July, he has been a standard expert with vivo Communication Research Institute. His research interests include system design and standardization for coding and modulation, MIMO, millimeter wave, artificial intelligence, etc.
\end{IEEEbiography}

\begin{IEEEbiography}[{\includegraphics[width=1in,height=1.25in,clip]{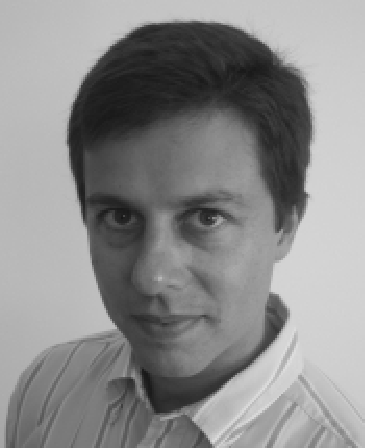}}]{Miguel Rodrigues}
(Senior Member, IEEE) received the Licenciatura degree in electrical and computer engineering from the University of Porto,
Porto, Portugal, and the Ph.D. degree in electronic and electrical engineering from the University College London (UCL), London, U.K. He is currently a Professor of Information Theory and Processing, UCL, and a Turing Fellow with the Alan Turing Institute - the UK National Institute of Data Science and Artificial Intelligence. His research lies in the general areas of information theory, information processing, and machine learning. His work has led to more than 200 articles in leading journals and conferences in the field, a book on Information-Theoretic Methods in Data Science (Cambridge Univ. Press), and the IEEE Communications and Information Theory Societies Joint Paper Award 2011. He is an Associate Editor for the IEEE TRANSACTIONS ON INFORMATION THEORY, and the IEEE OPEN JOURNAL OF THE COMMUNICATIONS SOCIETY. He was an Associate Editor for the IEEE COMMUNICATIONS LETTERS, and a Lead Guest Editor of the Special Issue on “Information-Theoretic Methods in Data Acquisition, Analysis, and Processing” of the IEEE JOURNAL ON SELECTED TOPICS IN SIGNAL PROCESSING. He was a Co-Chair of the Technical Programme Committee of the IEEE Information Theory Workshop 2016, Cambridge, U.K. He is a member of the IEEE Signal Processing Society Technical Committee on “Signal Processing Theory and Methods”, and the EURASIP SAT on Signal and Data Analytics for Machine Learning (SiG-DML).
\end{IEEEbiography}
%\end{comment}


%\vfill


\end{document}